\documentclass{aa}

\usepackage{graphicx}
\usepackage{txfonts}
\usepackage{mathrsfs}

\usepackage{epic}  
\usepackage{eepic}
\usepackage{pstricks}

\usepackage{amssymb}
\usepackage{amsmath}
\usepackage{textcomp}

\usepackage[]{natbib}
\usepackage{todonotes}
\usepackage{subfig}
\usepackage{booktabs}
\usepackage{placeins}

\def\Ps{\Pi}

\def\EE{\mathbb{E}}

\def\cH{{\cal H}}
\def\cL{{\cal L}}

\def\cF{{\cal F}}

\def\bxi{\mbox{\boldmath$\xi$}}

\def\rhat{\mbox{\boldmath$\hat{r}$}}
\def\zhat{\mbox{\boldmath$\hat{z}$}}

\def\thetahat{\mbox{\boldmath$\hat{\theta}$}}
\def\phihat{\mbox{\boldmath$\hat{\phi}$}}

\def\unit{\mbox{\boldmath$\hat{e}$}}

\def\ii{{\rm i}}
\def\id{{\rm d}}
\def\br{\mbox{\boldmath$r$}}

\def\xhat{\mbox{\boldmath$\hat{x}$}}

\def\zhat{\mbox{\boldmath$\hat{z}$}}

\def\bxi{\mbox{\boldmath$\xi$}}

\def\zhat{\mbox{\boldmath$\hat{z}$}}

\def\rhat{\mbox{\boldmath$\hat{r}$}}

\def\xhat{\mbox{\boldmath$\hat{x}$}}

\def\br{\mbox{\boldmath$r$}}

\def\one{\mbox{\boldmath$r$}_1}
\def\two{\mbox{\boldmath$r$}_2}
\def\bs{\mbox{\boldmath$r$}_{\rm s}}
\def\sbr{\mbox{\scriptsize\boldmath$r$}}

\def\sbu{\mbox{\scriptsize\boldmath$u$}}

\def\diff{\,\mathrm{d}}

\def\bF{\mbox{\boldmath$F$}}
\def\bf{\mbox{\boldmath$f$}}

\def\bu{\mbox{\boldmath$u$}}

\def\bxi{\mbox{\boldmath$\xi$}}

\def\unit{\mbox{\boldmath$e$}}

\def\rhat{\mbox{\boldmath$\hat{r}$}}
\def\thetahat{\mbox{\boldmath$\hat{\theta}$}}
\def\phihat{\mbox{\boldmath$\hat{\phi}$}}

\def\cG{{\mathcal G}}

\definecolor{amethyst}{rgb}{0.6,0.4,0.8}
\definecolor{Dorange}{rgb}{1,0.27,0}
\definecolor{forestgreen}{rgb}{0.13,0.54,0.13}

\title{Computational helioseismology in the frequency domain: acoustic waves in axisymmetric solar models with flows}
\titlerunning{Computational helioseismology}
\authorrunning{Gizon et al.}

   \author{
   Laurent Gizon
   \inst{1,2,}\thanks{Corresponding author: gizon@mps.mpg.de},
         H\'el\`ene~Barucq\inst{3,}\thanks{Corresponding author: helene.barucq@inria.fr}, 
Marc~Durufl\'e\inst{3},
Chris~S.~Hanson\inst{1}, 
Michael~Legu\`ebe\inst{1},
     Aaron~C.~Birch\inst{1},  
     Juliette~Chabassier\inst{3},
     Damien~Fournier\inst{4},
     Thorsten~Hohage\inst{4}, 
     Emanuele Papini\inst{1}
}
   \institute{
   Max-Planck-Institut f\"ur Sonnensystemforschung, Justus-von-Liebig-Weg 3, 37077 G\"ottingen, Germany
      	\and 
	Institut f\"ur Astrophysik, Georg-August-Universit\"at G\"ottingen, Friedrich-Hund-Platz 1, 37077      G\"ottingen, Germany
   \and 
	Magique-3D, INRIA Bordeaux Sud-Ouest, Universit\'e de Pau et des Pays de l'Adour, 64013 Pau, France
	\and
Institut f\"ur Numerische und Angewandte Mathematik, Georg-August-Universit\"at G\"ottingen, Lotzestra{\ss}e 16, 37083 G\"ottingen, Germany
	}

   \date{Received 15 June 2016; accepted XXX}

  \abstract
  {
 Local helioseismology has so far relied on semi-analytical methods to compute the spatial sensitivity of wave travel times to perturbations in the solar interior. These methods are cumbersome and lack flexibility.
 }
   {Here we propose a convenient framework for numerically solving the forward  problem of time-distance helioseismology in the frequency domain. The fundamental quantity to be computed is the cross-covariance of the seismic wavefield. 
   }
{
We choose sources of wave excitation that enable us to relate the cross-covariance  of the oscillations to the Green's function in a straightforward manner. 
We illustrate the method by considering the 3D acoustic wave equation in an axisymmetric reference solar model, ignoring the effects of gravity on the waves.
The symmetry of the background model around the rotation axis implies that the Green's function can be written as a sum of longitudinal Fourier modes, leading to a set of independent 2D problems. We use a high-order finite-element method to solve the 2D wave equation in frequency space. The computation is `embarrassingly parallel', with each frequency and each azimuthal order solved independently on a computer cluster. 
     }
   { We compute travel-time sensitivity kernels in spherical geometry for flows, sound speed, and density perturbations under the first Born approximation. Convergence tests show that travel times can be computed with a numerical precision better than one millisecond, as required by the most precise travel-time measurements. }
   {The method presented here is computationally efficient and will be used to interpret travel-time measurements in order to infer, e.g., the large-scale meridional flow in the solar convection zone. It allows the implementation of (full-waveform) iterative inversions, whereby the axisymmetric background model is updated at each iteration.
}


\begin{document} 

\maketitle

\section{Introduction}
Time-distance helioseismology and related techniques are methods for probing the complex dynamics of the solar convection zone   \citep{Duvall1993,GB2005, Gizon2010}. Information is encoded at the solar surface in the two-point cross-covariance function of the random solar oscillations. The cross-covariance function tells us about the travel time of wave packets between any two locations, in either direction. Flows break the time-symmetry of the cross-covariance function and thus leave a signature in the observations. 

Two topics of current interest include the study of meridional circulation \citep[e.g.,][]{Zhao2013,Liang_2015} and  convective flows \citep[e.g.,][]{Hanasoge2012, Langfellner2015} using, in particular, the SDO/HMI space observations. In both cases, small flow velocities are involved, resulting in travel-time perturbations that are less than a second. Hence it is not surprising that answers vary among investigators as they choose different interpretations and modeling procedures \cite[for a review see][]{Hanasoge2015}.

Helioseismic studies consist of several steps: measuring solar oscillations, processing and averaging the observations to extract the  seismic data (e.g., wave travel times), and interpreting the seismic data using forward and inverse methods to estimate solar internal properties. 
In this paper we mostly consider the forward problem, i.e. the computation of synthetic seismic data for a given solar model, which is  a necessary step for reliable interpretations. A short discussion of the iterative inverse problem is given at the end of the paper.

A framework was proposed by \cite{GB2002} to compute perturbations to the cross-covariance function caused by weak heterogeneities. This framework has proven to be useful for local helioseismology \citep{Birch2004, Birch2007, Birch2007b, Jackiewicz2008, Svanda2011, Hanasoge2011, Burston2015,Boening2016}. However, the computational expense has been a limiting factor in applications. Advances in the fields of earth seismology, exploration geophysics and ocean acoustics provide some guidance for improvements of the computational methods and the PDE-constrained formulations of the forward and inverse methods. 
\cite{Hanasoge2011} took a step in the direction of incorporating some of these ideas in the time domain within the framework of \cite{GB2002}. 
Of particular relevance to the present work is the proposal by \citet{NissenMeyer2014} to consider 3D wave propagation in relevant axisymmetric background media, in order to reduce the computational cost. Using a spectral decomposition in the azimuthal direction, the forward problem separates into many independent 2D problems, which can be solved to study the spatial sensitivity of seismic travel times to 3D heterogeneities \citep{vanDriel2014, vanDriel2015, Bottero2016}. In addition, theoretical studies have led to a better understanding of the connection between the cross-covariance function and the Green's function in media permeated by random sources of excitation \citep[][and references therein]{Snieder2013AR}.

In this paper we consider a number of current challenges for computational local helioseismology and propose some solutions.

\subsection{Challenges}

Modeling in local helioseismology faces a variety of challenges. Some of these challenges include:
\begin{enumerate}[C9.]
\item[C1.] 
Oscillation power spectrum. The purpose of helioseismology is to infer a model for the solar interior (density, sound speed, flows, etc., as functions of position) that is consistent with the seismic data.  In global mode helioseismology, the emphasis is on the comparison between the model and observed mode frequencies.  In contrast, local helioseismology requires models for the mode amplitudes and line profiles in addition to the mode frequencies. Thus a solar model must also describe wave excitation and damping. Most reference solar models are 1D and their quality can be assessed by comparing the model and observed oscillation power spectra.  In linear inversions for local helioseismology, the reference oscillation power spectrum has to be very solar-like.
\\
\item[C2.] Flexibility. The semi-analytical methods used so far to solve the forward problem \citep{GB2002} have proved useful \citep[e.g.][]{Burston2015} but lack flexibility. To first order (first Born approximation), the perturbation to the cross-covariance function is given by interaction integrals over products of Green's functions. Various approximations are made to speed up the computation of these integrals, for example using a local Cartesian geometry and neglecting some anisotropic effects. A generalization to more realistic setups, including the treatment of various instrumental (e.g., varying point-spread function) and geometrical effects (e.g., foreshortening, line-of-sight projection) is very cumbersome \citep[e.g.,][]{Jackiewicz2007}. How to include these complex effects in forward models in a reliable and efficient way is an open question. 
\\
\item[C3.] Representation of Green's function. Often Green's functions are computed using a normal-mode expansion \citep{Birch2004}. This approach is problematic for a  heterogeneous 3D reference model, for which eigenmodes are not readily available. Even if they were known, the summation formula would be computationally expensive. An additional complication is the treatment of the continuous spectrum above the acoustic cutoff frequency ($5.3$~mHz).
\\
\item[C4.] Time-domain simulations.
Alternatively, the Green's function may be computed numerically by solving the equations of motion in the time domain \citep[e.g.][]{hanasoge_duvall_2007, Cameron2008}. However, this requires a stabilization of the background model by changing the buoyancy frequency \citep[e.g.][]{Schunker2011, Papini2014}. Unless this operation is performed first, the linearized equations allow convective modes that grow exponentially. Unfortunately the mode frequencies are seriously affected by the stabilization of the solar model and become too far from the solar observations \citep{Papini2014}. 
\\
\item[C5.] Computational challenge and inverse problem.
Modern helioseismic observations consist of large data sets (long time series of $16$~megapixel images). Cross-correlations span a huge five-dimensional space. Extracting the relevant  information from this data set requires a good strategy and an efficient forward solver. This is especially true for iterative inversions, in which the forward solver is run for each update of the model of the solar interior.
\end{enumerate}

\subsection{Proposed approach}

We  propose to address the above issues by carrying out the following steps: 
\begin{enumerate}[C9.]
\item[1.] 
Rewrite the perturbation to the cross-covariance function
in terms of the Green's function, $G$, and the expectation value of the cross-covariance function, $\overline{C}$, in the reference model. In this formalism $G$ and $\overline{C}$ are the fundamental quantities that enter any forward calculation in local helioseismology. The problem then becomes deterministic. Additionally, many systematic effects may be accounted for by treating them as numerical operations on $G$ and $\overline{C}$.
\item[2.] 
Compute $G$ and $\overline{C}$ in the frequency-domain. The advantage of this approach is that frequencies are independent of each other for wave propagation in a steady background (many important problems in helioseismology fall in this category). This allows trivial parallelization of the computation in a multi-core environment. Furthermore, the computation can easily be restricted to the frequency range of interest, at the necessary frequency resolution.

\item[3.] 
Adopt a flexible geometrical set-up using a finite-element discretization of the problem. This will enable us to treat problems in which the required spatial resolution depends on both  the vertical and horizontal coordinates. Spherical geometry is naturally possible in this setup.
\item[4.] 
Consider a simplified problem of scalar acoustics in a stratified axially-symmetric reference model, for the sake of simplicity. Because of the axial symmetry the problem separates into many independent 2D problems, one for each azimuthal order. The parallelization in both frequency and azimuthal order allows for very efficient computation.
\item[5.]
Demonstrate that there exists a choice of source covariance, such that the expectation value of the cross-covariance is directly related to the Green's function, even in the presence of a background flow. In this setting, the Green's function is the only fundamental quantity that needs to be computed. 
\end{enumerate}

We will use the above approach to compute model power spectra, time-distance diagrams, and travel-time sensitivity kernels.
An application including a large-scale meridional flow will be presented. Though we will consider a scalar observable in this paper, the proposed approach is intended to be generalized to more realistic observables in the future.

\section{Pure acoustics in the frequency domain}

\subsection{Solar oscillations}

In an inertial frame, the displacement $\bxi(\br, t)$ of small amplitude waves obeys an equation of the form
\begin{equation}
\left(\frac{\partial}{\partial t} + \gamma \otimes  + \bu\cdot\nabla \right)^2 \bxi
 + \mathcal{H}[\bxi]  = \bf ,
 \label{eq.master}
\end{equation}
where $\mathcal{H}$ is a spatial operator and $\bu(\br)$ is a background flow. Wave attenuation is accounted for by temporal convolution ($\otimes$) with the function $\gamma(\br, t)$. The function $\bf(\br, t)$  represents forcing by turbulent convection and is thus a realization drawn from a random process. 
The spatial differential operator $\mathcal{H}$ is Hermitian for appropriate choices of the boundary condition and inner product \citep[see][]{LyndenBell1967}. An approximation of $\mathcal{H}$ that captures the physics of acoustic oscillations is:
 \begin{equation}
\mathcal{H}[\bxi] = - \frac{1}{\rho} \nabla(\rho c^2 \nabla\cdot\bxi) + \mbox{terms involving gravity} ,
\label{eq.Hmaster}
\end{equation}
where $\rho$ and $c$ are the density and sound speed in the reference model. Assuming that the medium is steady, the problem can be written in frequency space as
\begin{equation}
\label{eq.Hmaster2}
- (\omega +  \ii \,\widehat{\gamma} + \ii \bu\cdot\nabla)^2  \, \widehat{\bxi}
 + \mathcal{H}[\widehat{\bxi}]  = \widehat{\bf},
\end{equation}
using the Fourier convention
\begin{equation}
\widehat{\bxi}(\br,\omega)= \frac{1}{2\pi} \int_{-\infty}^{\infty} \bxi (\br, t) \, {\rm e}^{\ii \omega t} \,  \id t.
\end{equation}
If, in addition, the source of excitation is statistically stationary, then the frequency components $\widehat{\bf}(\br,\omega)$ are independent random variables and the frequency components of the wave displacement, $\widehat{\bxi}(\br,\omega)$, are also statistically  independent of each other. Thus, in frequency space, the problem separates into a set of independent boundary-value problems, one for each frequency. 

In the following sections we drop the hat on top of Fourier transforms to simplify the notation.

\subsection{Scalar wave equation}

Let us further specify the wave equation to be solved.
There are two possible choices for reducing the equation for the wave displacement to a scalar equation. One possibility is to rewrite Eq.~\eqref{eq.Hmaster2} in terms of the quantity $\psi = c \nabla\cdot\bxi$. 
Neglecting  the gravity terms, the second-order terms in $\gamma$ and $\bu$, and the cross term in $\gamma \bu$ in Eq.~(\ref{eq.Hmaster2}), then taking the divergence, we obtain 
\begin{equation}
\begin{aligned}
& -(\omega^2 + 2\ii \omega \gamma) \psi  
- 2 \ii  \omega   c \, \nabla \cdot \left( \bu \cdot \nabla \bxi \right)
- 2\ii \omega c \, \bxi\cdot \nabla \gamma  \\
 & - c \nabla \cdot \left(  \frac{1}{\rho} \nabla \left( \rho c \psi \right) \right)
 = c \nabla \cdot \bf .
 \end{aligned}
 \end{equation}
 For slow variations of $\bu$, $c$ and $\gamma$ compared to the wavelength, the wave equation for $\psi$ simplifies to  
\begin{equation}
L[\psi] := - \sigma^2  \psi   - 2\ii\omega  \,  \bu\cdot\nabla  \psi
 +     H[\psi] = s 
\label{eq.scalar}
\end{equation}
with
\begin{equation}
\begin{aligned} 
&\sigma^2   :=  \omega^2 + 2\ii\omega\gamma 
\\ &
H[\psi] := - c \nabla\cdot \left( \frac{1}{\rho} \nabla (\rho c \psi) \right).
\end{aligned}
 \end{equation}
The random source of excitation,  $s(\br,\omega)=c(\br) \nabla \cdot \bf(\br,\omega)$, and the attenuation, $\gamma(\br, \omega) >0$, depend on frequency. 

Another possible choice to obtain a scalar equation is to use the ansatz $\bxi = \rho^{-1}  \nabla \left(  \rho c \psi \right)$. 
Under the assumption that the forcing term $\rho \bf$ is curl free,  this leads to the same scalar equation as above.

When $\rho$ and $c$ are solar-like, the scalar  equation \eqref{eq.scalar} captures most of the interesting physics of p modes.  For example, a solar-like density gives the correct acoustic cut-off frequency of $5.3$~mHz. However, this equation is not relevant for modeling f and g modes.

Here, we choose to specify the source of excitation through the covariance function
\begin{equation}
\label{eq.src}
M(\br,\br',\omega) = 
\EE [ s^*(\br, \omega) s(\br', \omega)] 
= \epsilon( \br)  \Ps(\omega) \delta(\br-\br')  ,
\end{equation}
where $*$ takes the complex conjugate and the functions $\epsilon(\br)$ and  $\Ps(\omega)$ control the spatial and 
frequency dependencies of the source covariance respectively. In the above expression   we assumed spatially uncorrelated sources. This formulation is standard in local helioseismology \citep[e.g.][]{Birch2004}.

\subsection{Boundary condition}

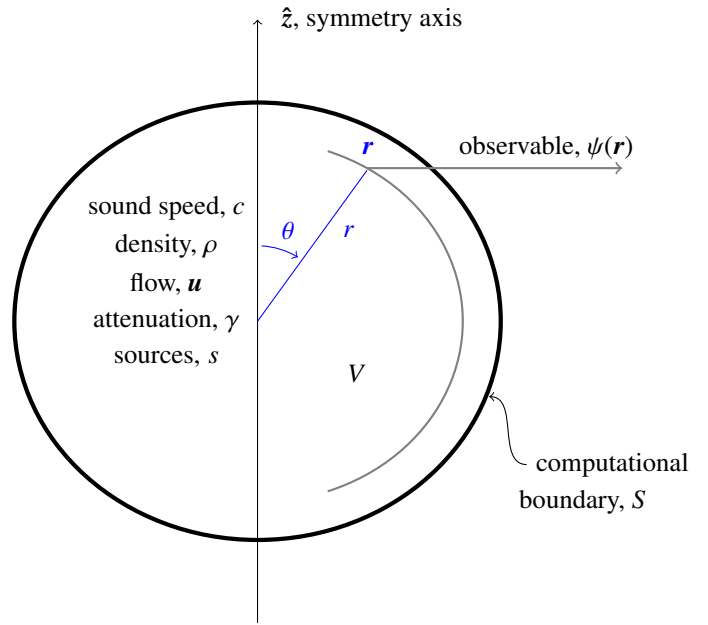
\begin{figure}[t]
\setlength{\unitlength}{1.2mm}
\begin{tikzpicture}
\draw[ultra thick](3.2,0) ellipse (3.2 and 2.9);
\draw[gray, thick]  (-70:2.7 and 2.4) ++(3.2,0) arc (-70:70:2.7 and 2.4);
\draw[->] (4*0.8,-4) -- (4*0.8,4);
\node[align=left] at (4.7,4) {$\zhat$, symmetry axis};
\draw[blue] (3.2,0.) -- (5.8*0.8,2.);
\draw[gray, thick, ->] (5.8*0.8,2.03) -- (10*0.8,2.03);
\node at (7,2.3) {observable, $\psi(\br)$};
\node[blue] at (4.65,2.3) {$\br$};
\node[align=left] at (2,1.5){sound speed, $c$};
\node[align=left] at (2,1.){density, $\rho$};
\node[align=left] at (2,0.5){flow, $\bu$};
\node[align=left] at (2,0.){attenuation, $\gamma$};
\node[align=left] at (2,-0.5){sources, $s$};

  \node[anchor=east] at (9,-1.9) (text) {computational};
  \node[anchor=west] at (6,-1) (description) {};
 \draw (text) edge[out=180,in=0,->] (description);
\node at (7.5,-2.4) {boundary, $S$};
\node at (4.5,-0.7) {$V$};
\node[blue] at (3.6,1.2) {$\theta$};
\draw[blue, <-] (58:1.)++(3.2,0) arc (58:87:1.);
\node [blue] at (4.4,1.2) {$r$};
\end{tikzpicture}
\caption{Geometrical setup for a background medium  symmetric about an axis $\zhat$. The thick contour delineates the boundary, $S$, of the computational domain $V$. A position vector $\br$ in $V$ is specified by spherical-polar coordinates  ($r$, $\theta$, $\phi$). The seismic observable $\psi(\br)$ is measured near the photosphere.
}
\label{fig:geometry}
\end{figure}

The geometrical set-up is shown in Fig.~\ref{fig:geometry}.
We work in a spherical-polar coordinate system $(r,\theta,\phi)$. The solar (or stellar) model is cylindrically symmetric around the $\zhat$ axis. The computational boundary $S$ encloses the solar photosphere. The wave equation is supplemented by a boundary condition at $S$ on $\psi$.  {We denote by $\mathbf{\hat{n}}$ the unit normal vector to $S$ and by $\partial_n \psi$ the normal derivative of $\psi$.}

A free-surface boundary condition ($\psi\propto \nabla\cdot\bxi = 0$, Dirichlet) is often used in helioseismology \citep[][etc.]{bogdan_etal_1996,JCD}. However, it is more appropriate to choose a transparent boundary condition to model the 
continuous spectrum above $5.3$~mHz
and to allow high-frequency waves to escape \citep[e.g.][]{Kumar1990}. The use of perfectly matched layers is one way to approximate a transparent boundary \citep[e.g.][]{Hanasoge2011}. Here, we choose the simplest approximation to the  Sommerfeld outgoing radiation condition, 
\begin{equation}\label{eq.sommerfeldBC}
\partial_n \psi  =  \ii  k_n  \psi  \quad \mbox{ on } S ,
\end{equation}
where $k_n$ is the local wavenumber normal to the boundary.
In order for this radiation boundary condition to hold true, the sound speed and density must be constant near the computational boundary, which is not the case in the Sun. In practice, we extend the solar reference model with a layer that smoothly transitions to constant density and sound speed (see Sect.~\ref{sec.smoothModelS} for details).

\subsection{Hermiticity}\label{sec.hermiticity}

In this section and the following, we demonstrate some important properties of the scalar wave equation~\eqref{eq.scalar}, which will be used in the later sections. These basic properties would remain the same in the full vector equations.

One important property of the equations that describe non-attenuating waves  in the Sun is the Hermiticity of the wave operator, which implies that the eigenvalues are real. 
\citet{LyndenBell1967} proved Hermiticity of the wave operator for a self-gravitating fluid bounded by a free surface. 

Here we ask if this condition is satisfied for the time-harmonic scalar wave equation defined by Eq.~\eqref{eq.scalar}. The scalar wave field belongs to a complex Hilbert space. For any two functions $\psi$ and $\varphi$ in this space we define the inner product
\begin{equation}
\langle \psi , \varphi  \rangle = \int_V \psi^* \varphi \, \rho \id\br  ,
\label{eq.inner}
\end{equation}
where $V$ is the computational domain. Performing integration by parts twice implies
\begin{equation}
\label{eq.self-H}
\langle \psi , H[\varphi]  \rangle = \langle H[\psi] ,  \varphi  \rangle  
-  \int_S  \left[ \psi^*  \,  \partial_n ( \rho c \varphi )   -      \varphi   \, \partial_n (\rho c \psi^*)  \right]   c \diff S.
\end{equation}
We see that the spatial operator $H$ is Hermitian if $\psi$ and $\varphi$  vanish on the boundary. However, for a radiation boundary condition,  the surface integral remains and $H$ is not Hermitian.
The eigenvalues of $H$ are no longer real. 

The other terms in equation~\eqref{eq.scalar} are the damping and advection terms. Using integration by parts once, we
ascertain that the damping operator is anti-Hermitian and the advection operator is Hermitian: 
\begin{equation}
\label{eq.self-gamma}
\langle \psi, 2\ii\omega \gamma\varphi  \rangle=
 - \langle 2\ii\omega \gamma \psi , \varphi \rangle ,
\end{equation}
\begin{equation}
\label{eq.self-flow}
\langle \psi, 2\ii\omega \bu\cdot\nabla\varphi  \rangle=
\langle 2\ii\omega \bu\cdot\nabla\psi , \varphi \rangle .
\end{equation}
The last result is a consequence of mass conservation, $\nabla\cdot (\rho \bu) =0$, for a steady flow that does not cross the computational boundary 
{($\bu \cdot \mathbf{\hat{n}} = 0$ on $S$).} 

Combining Eqs.~\eqref{eq.scalar}, \eqref{eq.self-H}, \eqref{eq.self-gamma}, and \eqref{eq.self-flow} yields 
\begin{equation}
\label{eq.Ldag}
\langle \psi , L[\varphi]  \rangle = \langle L^\dagger [\psi] ,  \varphi  \rangle -  \int_S   \left[ \psi^*    \partial_n \left( \rho c \varphi \right)   -    \varphi  \partial_n (\rho c \psi^*)    \right]     c   \diff S ,
\end{equation}
where  
\begin{equation}
L^\dagger = L_{(-\gamma)} = L^*_{(-\sbu)} 
\end{equation} 
 is the operator obtained by switching the sign of $\gamma$ in $L$ or, equivalently, by switching the sign of  $\bu$  and taking the complex conjugate.  
 Note that for a radiation boundary condition, $L^\dagger$ is not the adjoint of $L$ because  the surface integral remains.

\subsection{Green's function and generalized seismic reciprocity}
In geophysics, the principle of seismic reciprocity states that the same signal should be recorded if the locations of a source and a receiver are exchanged. 
Although seismic reciprocity is not preserved in the presence of a flow, 
a generalization of reciprocity can be obtained. 

Let us introduce the Green's function as the solution to
\begin{equation}
\label{eq.green}
 L [G (\br,\br', \omega)]     =  \frac{1}{\rho(\br)} \delta(\br -\br') ,
 \end{equation}
 where $\delta$ is the Dirac delta function and $G$ satisfies the same boundary condition as the  wave field $\psi$. Consider a source at $\one$ in the physical domain and a source at $\two$ in the domain in which the flow has opposite sign:
\begin{equation}
\left\{
\begin{aligned}
& \rho \, L [G (\br,\one, \omega)]     =  \delta(\br -\one)   \\
& \rho \, L_{(-\sbu)}[G (\br,\two, \omega; -\bu)]   =  \delta(\br-\two)  ,
\end{aligned}
\right.
\end{equation}
where $G(\cdot; -\bu)$ is the Green's function associated with the operator $L_{(-\sbu)}$, with the same boundary condition as $G$.
Multiplying the first equation by $G(\br,\two, \omega; -\bu)$ and the second by $G(\br,\one, \omega)$, integrating each over position $\br$, and subtracting the two equations, we obtain 
\begin{equation}
\label{eq.GLG}
\begin{aligned}
&G(\br_1,\br_2, \omega; -\bu) - G(\br_2,\br_1, \omega) = \\  & \qquad
   \left\langle    G^*(\br, \br_2, \omega; -\bu) ,  L [G(\br,\br_1, \omega)] \right\rangle \\
& \qquad - \langle  L^\dagger [G^*(\br,\br_2, \omega;-\bu)] ,  G(\br, \br_1, \omega)  \rangle .
\end{aligned}
\end{equation}
Following the same logic as earlier (integration by parts),  
\begin{equation}
\langle \psi^*,L[\varphi]\rangle = \langle L^\dagger[\psi^*],\varphi\rangle -
\int_S \left[ \psi\partial_n\left(\rho c\varphi\right) 
- \varphi\partial_n\left(\rho c\psi\right)\right] c \id S
\end{equation}
for  any two functions  $\psi$ and $\varphi$ in the space of solutions. This time the surface integral vanishes for either type of boundary condition (Dirichlet or radiation). In particular, 
 \begin{equation}
\langle   L^\dagger   [G^*], \varphi \rangle  = \left\langle  G^* , L [\varphi]  \right\rangle.
\end{equation}
This relationship shows that the right-hand side of Eq.~\eqref{eq.GLG} vanishes and so we find
\begin{equation}
G(\two,\one, \omega) =    G(\one,\two, \omega; -\bu)   .
 \end{equation}
 This is a generalization of seismic reciprocity:  the Green's function is unchanged upon exchanging  source and receiver and changing the sign of the flow. The conclusion holds even though the damping operator is not Hermitian.

\section{Cross-covariance function and travel times}

For the sake of simplicity, let us suppose that $\psi$ can be directly measured on the solar surface.  This choice is not unreasonable as  $\psi$ is proportional to the pressure fluctuations and is thus related to intensity fluctuations, an observable quantity.
This choice is not at all a limitation of the method: other observables can be derived from $\psi$, including for example a proxy for the line-of-sight velocity.

At frequency $\omega$, consider the cross-covariance in Fourier space as the product of the wave field at two locations of measurement, 
\begin{equation}
\begin{aligned}
&C(\one, \two,\omega) =  \psi^*(\one,\omega)  \, \psi(\two,\omega)   .
 \end{aligned}
\end{equation}
In terms of the Green's function, we have
\begin{equation}
\psi (\br_j, \omega)  = \int_V G(\br_j, \br, \omega)  s(\br, \omega) \, \rho \id\br   , 
\end{equation}
where the source $s(\br, \omega)$ is a realization of a random process.
Under the assumption of spatially uncorrelated sources (see Eq.~\eqref{eq.src}),
the cross-covariance can be written as a single volume integral:
\begin{equation}
\label{eq.EC}
\begin{aligned}
&\overline{C} (\one, \two,\omega)  = \EE[  C(\one, \two, \omega) ] 
\\
& \qquad =
\int_V \rho \id\br \int_{V} \rho'\id \br' \; G^*(\one, \br, \omega)  \, G(\two, \br', \omega)  \, M(\br,\br', \omega)       
\\
& \qquad = \Ps(\omega) \int_V     G^*(\one,\br, \omega)   \, G(\two,\br, \omega)  \,  \epsilon(\br)  \, \rho^2 \id\br  .
\end{aligned}
\end{equation}

Following \citet{GB2002}, we define the perturbation to the travel time $\delta \tau$ between points $\one$ and $\two$ as
\begin{equation}
\label{eq:dtCross}
\delta \tau (\one, \two)  =    \int_{-\infty}^\infty  W^*(\omega)  \, [ C(\one,\two,\omega) - C_{\rm ref} (\one,\two,\omega) ]  \diff\omega  ,
\end{equation}
where
\begin{equation}
 W(\omega) = - \frac{  \int_{-\infty}^\infty   w(t) \, [\partial_t C_{\rm ref}(\one,\two,t)] \, {\rm e}^{\ii\omega t} \, \id t }{\int_{-\infty}^\infty   w(t')\, [\partial_{t'} C_{\rm ref}(\one,\two,t')]^2 \, \id t' } ,
 \end{equation}
$C_{\rm ref}$ is a reference cross-covariance function, and $w(t)$ is a temporal window function that selects a particular section of the data. For example, we may choose $w(t) = \exp[-(t-t_{\rm g})^2/2\sigma^2]$ where $t_{\rm g}>0$ is a group time and $\sigma$ sets the width of the window. 
 
\section{New formulation of travel time sensitivity kernels}
\label{sec.kernelForm}

In the presence of subsurface perturbations to the reference model the travel times computed from Eq.~\eqref{eq:dtCross} will be affected. An understanding of the spatial sensitivity of the measurements to structural perturbations and flows requires the development of travel time sensitivity kernels. In this section we develop a new formulation of the travel-time sensitivity kernels in terms of the Green's function and the cross-covariance computed in the reference model.
 
\subsection{Perturbations to the medium}
In addressing how travel times are sensitive to changes in the solar reference model, we will consider changes in $c(\br)$, $\rho(\br)$ and $\bu(\br)$, as well as spatial changes to the attenuation $\gamma(\br)$ and the source amplitude $\epsilon(\br)$.
The vector flow 
\begin{equation}
\bu(\br) = \sum_k u_k(\br) \, \mathbf{\hat{e}}_k(\br) = u_r(\br) \rhat  +  u_\theta(\br)  \thetahat +  u_\phi(\br) \phihat , 
\end{equation}
is specified by the three components $u_k$ on the basis $\{\mathbf{\hat{e}}_k\}$ of  unit vectors $\rhat$, $\thetahat$ and $\phihat$ for the spherical-polar coordinate system.
In short, the physical variables of interest may be combined into a set
\begin{equation}
\{ q_\alpha   \} = \{ c, \rho , u_r,u_\theta,u_\phi, \gamma, \epsilon \}.
\end{equation}
We are looking for travel-time sensitivity kernels $K_\alpha$ such that 
the travel-time perturbation $\delta \tau$ can be written as 
\begin{equation}
\EE [ \delta\tau ]= \sum_\alpha \int_V  \delta q_\alpha (\br) K_\alpha (\br)  \diff\br  
+ \mbox{ surface terms}    ,
\label{eq.kerdef}
\end{equation}
for infinitesimal perturbations $\delta q_\alpha$. Note that for some variables there is a surface integral on the computational boundary, because integration by parts is needed to write travel-time perturbations in the form of Eq.~\eqref{eq.kerdef}. For these variables, we assume that the perturbations to the model vanish on the computational boundary (high in the atmosphere) and we drop the surface terms. 

In order to find the kernels, the travel-time perturbation
\begin{equation}
\label{eq.taudC}
\delta\tau (\one,\two) =  \int_{-\infty}^\infty  W^*(\omega)  \, \delta C(\one, \two, \omega)\diff\omega 
 \end{equation}
is written in terms of the first-order perturbation to the cross-covariance:
 \begin{equation}
  \delta C(\one,\two, \omega) = 
  \psi^*(\one, \omega) \, \delta \psi(\two, \omega ) +   
  \delta \psi^*(\one, \omega)  \, \psi(\two, \omega ) .
  \label{eq.dccc}
  \end{equation}
The next step is then to express  $\delta\psi$ in terms of the $\delta q_\alpha$.  
 
\subsection{Perturbation to the wave field}

To first order (first Born approximation) the operator defined in Eq.~\eqref{eq.scalar} acts on the perturbed wave field through,
\begin{eqnarray}
  L [\delta\psi]  & = &  - \delta L[\psi]  + \delta s
\nonumber   \\ &=  &  
2\ii\omega \, \delta \gamma \, \psi
+ 2\ii \omega \, \delta\bu\cdot\nabla \psi
 - \delta H [\psi]   + \delta s ,
 \label{eq.born}
  \end{eqnarray}
where $\delta L$ is the perturbation to the wave operator caused by perturbations to the medium and
\begin{eqnarray}
 \delta H [\psi] 
  &=&  \frac{\delta c}{c} H [\psi]  + H \left[\frac{\delta c}{c}\psi \right] 
  \nonumber \\ &&  
  - c \nabla\cdot \left(   \frac{\delta \rho }{\rho^2} \nabla (\rho c \psi) \right) +  H \left[\frac{ \delta \rho}{\rho}   \psi   \right]. 
\end{eqnarray}
A formal solution to Eq.~\eqref{eq.born} is obtained in terms of the  Green's function: 
\begin{equation}\label{eq.pertwave}
\begin{aligned}
\delta \psi(\br_j, \omega)   =&  -  \int_V   G(\br_j,\br, \omega) \, \delta L[ \psi (\br, \omega)] \,  \rho\id \br \\
& + \int_V G(\br_j,\br, \omega)  \, \delta s(\br, \omega)\, \rho \id\br .
\end{aligned}
\end{equation}

\subsection{Perturbation to the cross-covariance}
With Eq.~\eqref{eq.pertwave} in hand, the expectation of the perturbation to the  cross-covariance is then determined from  Eq.~\eqref{eq.dccc}:
\begin{equation}
\begin{aligned}
\label{eq.dC1}
\delta\overline{C}(\one,\two, \omega)  = & -    \int_V   G(\two,\br, \omega) \; \delta L\left[ \overline{C}(\one, \br, \omega)\right] \,  \rho \id \br 
\\&
 -   \int_V   G^*(\one,\br, \omega)  \; \delta L^*\left[   \overline{C}^{\, *}\!(\two, \br, \omega) \right] \,  \rho \id \br 
\\&
+   \Ps(\omega) \int_V G^*(\one, \br, \omega) G(\two,\br, \omega)  \, \delta\epsilon (\br) \, \rho^2 \id \br ,
\end{aligned}
\end{equation}
where we write the perturbation to the source covariance as a   spatially-varying change in  amplitude,
\begin{equation}
 \EE[  s^*(\br, \omega) \delta s (\br', \omega) +   \delta s^*(\br, \omega)  s (\br', \omega)  ] 
 = \delta \epsilon( \br)  \Ps(\omega) \delta (\br -\br') .
 \end{equation}
In order to obtain kernels for individual perturbations, we introduce the  bilinear operators $\cL_\alpha$ such that for any functions $g$ and $h$ we have
\begin{equation}
\label{eq.Lalpha}
\int_V g \; \delta L [h]  \, \rho \id\br  =   \sum_\alpha \int_V  \delta q_\alpha(\br) \,  \cL_\alpha [ g, h ]\,  \id\br  + \mbox{surface terms} .
\end{equation}
These bilinear operators are obtained by integration by parts of the above left-hand side  and are explicitly given by
\begin{eqnarray}
\cL_c [ g, h  ] &=&      \left( g \, H [h]  +      h\, H [g] \right) \rho/c ,
   \\
 \cL_\rho [ g, h ]  &=&     - \nabla(\rho c g)   \cdot  \nabla(\rho c h)   / \rho^2 +   h\, H [g]         ,
\\
 \cL_{u_k} [ g, h  ]  &=&     - 2 \ii \rho \omega \,    g  \, (\partial_k h)   , \label{eq.cLu}
  \\ 
 \cL_\gamma  [g, h ] &=& -  {2 \ii \rho\omega }  \, g \,    h  .
\end{eqnarray}
 In Eq.~\eqref{eq.cLu}, $\partial_k = \hat{\unit}_k(\br) \cdot\nabla_{\sbr}$ denotes the component of the  spatial gradient in the $k$ direction. 
 Combining Eqs.~(\ref{eq.Lalpha}) and (\ref{eq.dC1}),  we obtain an explicit linear relationship between the $\delta q_\alpha$  and $\delta\overline{C}$:
\begin{equation}
\begin{aligned}
&\delta\overline{C}(\one,\two,  \omega)  = \\
 & =   - \sum_\alpha \int_V   \delta q_\alpha(\br)  \; \cL_\alpha\left[ G(\two,\br,\omega),  \overline{C}(\one,\br,\omega)\right]  \id \br 
\\
& \quad - \sum_\alpha \int_V   \delta q_\alpha(\br)  \;     \cL_\alpha^*\left[ G^*(\one,\br,\omega),  \overline{C}^{\,*}\!(\two,\br,\omega)\right]  \id \br 
\\
& \quad +   \Pi (\omega) \int_V   \delta\epsilon (\br)  \, G^*(\one,\br, \omega) G(\two,\br, \omega) \, \rho^2 \id\br 
\\
& \quad + \text{ surface terms}.
\end{aligned}
\end{equation}
Using this expression together with Eqs.~\eqref{eq.kerdef} and (\ref{eq.taudC}), the travel-time sensitivity kernels  are
\begin{equation}
\begin{aligned}
K_\alpha (\br ; \one, \two)  = &   -  \int_{-\infty}^\infty   W^*(\omega) \; \cL_\alpha \left[ G (\two,\br,\omega)  ,   \overline{C}(\one,\br,\omega)  \right] \id\omega  
\\ & 
-   \int_{-\infty}^\infty  W^*(\omega) \;    \cL_\alpha^* \left[ G^*(\one,\br, \omega) ,   \overline{C}^{\, *}\!(\two,\br, \omega)  \right] \id\omega  
   \end{aligned} \label{eq.kernelGeneral}
   \end{equation}
   for the scatterers $\delta q_\alpha\in
\{ \delta c,  \delta \rho,  \delta u_k, \delta \gamma \}$
   and 
   \begin{equation}
 K_\epsilon (\br ; \one, \two)  =    \rho^2(\br) \int_{-\infty}^\infty    \Ps(\omega)   W^*(\omega) \,    G^*(\one,\br, \omega)   G (\two,\br, \omega)  \, \id \omega
\end{equation}
for the source amplitude $\delta \epsilon$.

\section{Convenient source of excitation}\label{sec.convSource}

\subsection{How can we simplify the computation of $\overline{C}$?}

The kernels from the previous section are well defined once the background medium, attenuation and source functions have been specified. At each $\omega$, the attenuation function $\gamma(\br,\omega)$ could be tuned to yield the observed line widths of solar oscillations in the power spectrum. Similarly, at each $\omega$, the source function $\epsilon(\br)\Pi(\omega)$  could be tuned to give the observed mode amplitudes \citep{Birch2004, Birch2007}. However, the integral relationship between a general $\epsilon(\br)$ and $\overline{C}$ is far from trivial to compute.

Here we adopt a different strategy. 
We ask whether there exists a convenient source covariance such that the expectation value of the cross-covariance can be written in terms of the Green's function. As a second step we will check whether the resulting power spectrum is solar-like.

It is known in geophysics and acoustics that, under appropriate conditions on the source covariance, the expectation value of the cross-covariance in the frequency domain is related to the imaginary part of the Green's function \citep[see][and references herein]{snieder_etal_2009}.  If we could write such a simple relationship, our problem would simplify considerably. For all practical purposes, the Green's function would be the only truly important quantity.

Let us start by rewriting the equation for the Green's function (Eq.~\ref{eq.green}) for sources at $\one$ and $\two$, taking the complex conjugate of the first:
\begin{equation}
\left\{
\begin{aligned}
&\delta(\br -\one)  = \rho(\br) \, L^*    [G^*(\br,\one, \omega)]        , 
\\
&\delta(\br-\two) = \rho(\br)  \, L   [G(\br,\two, \omega)]       .
\end{aligned}
\right.
\end{equation}
Multiply the first equation by $G(\br,\two, \omega)$ and the second equation by $G^*(\br,\one, \omega)$,   integrate each over $\br$, and then subtract  to find
\begin{equation}
\begin{aligned}
&  G(\one,\two, \omega) - G^*(\two,\one, \omega)  = 
\\ 
& = \langle L [G(\br,\one, \omega)]  , G(\br, \two, \omega)  \rangle - \langle G(\br, \one, \omega) , L [G(\br,\two, \omega)] \rangle 
\\  
&  = \langle L [G(\br,\one, \omega)]  , G(\br, \two, \omega)  \rangle - \langle L^\dagger [G(\br, \one, \omega)] ,  G(\br,\two, \omega) \rangle  \\
&\quad +   \mbox{surface term} 
\\  
&  = 4\ii \omega  \langle \gamma(\br, \omega)  G(\br,\one, \omega)  , G(\br, \two, \omega)  \rangle +  \mbox{surface term} 
\\  
&  = 4\ii \omega  \langle \gamma(\br, \omega) G(\one,\br, \omega; -\bu)  , G(\two, \br, \omega; -\bu)  \rangle +  \mbox{surface term.}
\end{aligned}
\end{equation}
The surface term does not vanish unless specific boundary conditions are used, as noticed earlier by  \citet{Snieder2007a}. In the solar  case (transparent boundary condition), this surface term does not vanish.  Reversing the flow $\bu$ in the above equation, we obtain
\begin{equation}
\begin{aligned}
& G(\two, \one, \omega) - G^*(\two,\one, \omega; -\bu) = 
\\
 &  = 4\ii\omega \int_V   \gamma(\br,\omega)     G^*(\one,\br, \omega)   G(\two, \br, \omega)\, \rho \id V 
 \\ & \quad
  + 2\ii \omega \int_S    c(\br)  G^*(\one,\br,\omega)   G(\two, \br,\omega) \, \rho \id S  ,
 \end{aligned}
 \end{equation}
 where the surface integral is the explicit expression for the surface term. By inspection of Eq.~\eqref{eq.EC}, we see that the choice of source covariance amplitude
\begin{equation}
\label{eq.simplesource}
 \epsilon :=  \frac{\gamma(\br,\omega)}{\rho(\br)}  + \frac{c}{2\rho}  \delta ( r   -  R(\theta)  )
\end{equation}
implies
\begin{equation}
\label{eq.EC-ImG} 
\EE[ C(\one,\two, \omega) ]=   \frac{\Ps(\omega)}{4\ii\omega}  \left[  G(\two,\one, \omega)  - G^*(\two,\one, \omega; -\bu) \right] .
\end{equation}
Thus the expectation value of the cross-correlation can be written as a sum of causal and anti-causal Green's functions, when waves are appropriately excited throughout the volume and on the computational  boundary $S$ at radius $r=R(\theta)$. The volume sources must be proportional to the local attenuation to enforce energy equipartition between the modes \citep[see, e.g.,][]{Snieder2007a}.

To simplify the problem it would be very nice to assume that equality~\eqref{eq.EC-ImG} holds. Does it give a reasonable oscillation power spectrum? The power spectrum can be tuned in frequency space by choosing $\Ps(\omega)$. However, the source of Eq.~\eqref{eq.simplesource} leaves no freedom for the distribution of power versus radial order at fixed frequency.

A drawback of the choice of source (Eq.~\eqref{eq.simplesource}) is that the source covariance cannot be updated independently from attenuation and density. Thus, perturbations to the cross-covariance are fully specified  by perturbations to $c$, $\rho$, $\gamma$, and $\bu$. 

\subsection{Sensitivity kernels in terms of $G$ only}

As the cross-covariance depends only on the Green's function via Eq.~\eqref{eq.EC-ImG}, it is also the case for the kernels. Denoting for simplicity 
\begin{equation}
\begin{aligned}
&G_i (\br) :=  G(\br, \br_i, \omega),  \\
& G^\dagger_i (\br) :=  G^*(\br, \br_i, \omega ; -\bu), 
\end{aligned}
\end{equation}
the kernels defined by Eq.~\eqref{eq.kernelGeneral} can be computed from four Green's functions
\begin{equation}
G_1(\br),  \: G_2(\br) \quad {\rm and} \quad  G_1^\dagger(\br),  \: G^\dagger_2(\br).
\end{equation}
For example, the travel-time perturbations induced by  a background flow perturbation,
\begin{equation}
 \delta \tau ( \one,\two) = \sum_k \int_V \delta u_k(\br) \,  K_{u_k}(\br;\one,\two)   \, \id\br   ,
\end{equation}
can be computed through the kernels
\begin{equation}
K_{u_k}    =     \frac{\rho}{2} \int_{-\infty}^\infty \Ps  W^*   \left[  G^{\dagger*}_2  \partial_k \left( G_1  - G^\dagger_1 \right)  + G^\dagger_1   \partial_k  \left( G_2^*  - G^{\dagger *}_2  \right)   \right] \, \id \omega .
\end{equation}

\subsection{Special case of no background flow}
Following from the previous section and taking the further assumption of no background flow ($\bu_0=0$), seismic reciprocity implies that $G^\dagger = G^*$ and the cross-covariance takes the simple form,
\begin{equation}
\overline{C}_i(\br) := \overline{C} (\br_i, \br, \omega ) =   \frac{\Ps(\omega)}{2\omega}  
\mbox{ Im }  G_i  (\br).
\end{equation}
This simple form for the cross-covariance simplifies the scattering bilinear operators:
\begin{eqnarray} \label{eq.bilinearFormKernel}
\cL_c [ G_i, \overline{C}_j  ] &=&        \frac{ \rho \Ps}{2\omega c}
\left[ G_i \,    {\rm Im} ( H[G_j])  +    H[G_i]  \, {\rm Im}  G_j \right]  ,  \\
 \cL_\rho [ G_i, \overline{C}_j ]  &=& \!\!\!\!  -  \frac{\Ps }{2\omega }\left(  \frac{1}{\rho^2} \nabla(\rho c G_i)   \cdot   \nabla(\rho c  \, {\rm Im} G_j)   -   H[G_i]  \,   {\rm Im} G_j     \right) ,\nonumber
\\ &&\\
 \cL_{u_k} [ G_i, \overline{C}_j  ]  &=&   -    \ii \rho \Ps  \,    G_i  \,       \partial_k ({\rm Im} G_j) ,
  \\ 
 \cL_\gamma  [G_i, \overline{C}_j ] &=&   - \ii  \rho  \Ps \, G_i   \, {\rm Im} G_j .
 \label{eq.bilinearFormKernel.end}
\end{eqnarray}
Thus, to compute a kernel $K_{\alpha}$ when there is no background flow, all we need is two Green's function  computations, one  with a source at $\one$ and the other with a source at $\two$.

\section{Forward solver in the frequency domain}

With the theory of the previous sections in hand, we wish to compute the Green's function at fixed $\omega$, defined by Eq.~\eqref{eq.green}. 
In order to achieve this we utilize a standard Finite Element Method (FEM), {which is further explained in Sect.~\ref{sec.weakform}.}

Ideally we would solve for the Green's function in a general 3D computational domain. 
However, as discussed later in Sect.~\ref{sec.comptimes}, 3D computations are very demanding.
In the following subsections we consider a reference  model that is symmetric about an axis. 
This set up is appropriate to study the effects of large-scale flows (differential rotation and meridional flow) on solar oscillations. 
A substantial benefit of this choice is that the solution to the 3D problem can be obtained by decomposition into a set of independent 2D problems, one for each Fourier component in longitude. 
We refer to problems where the background medium is axisymmetric as $2.5$D problems. Below we describe the computational setup and give a weak formulation of the solved equations.

\subsection{Montjoie finite element code}

We utilize the FEM code Montjoie, developed at Inria and described at \texttt{http://montjoie.gforge.inria.fr/}, which was originally developed for various wave propagation problems. 
Montjoie is versatile, well-tested, and robust \citep[see][]{duruflethesis,bergot10}. 
Montjoie allows us to consider  $1.5$D, $2.5$D, and 3D geometries. 
The $1.5$D problem assumes a spherically symmetric background model and the solution is decomposed into  spherical harmonics and 1D finite radial elements. 
The $2.5$D method assumes an axisymmetric background model and the solution is decomposed into azimuthal Fourier modes and 2D finite elements; it is fast and useful for a number of applications. 
The full 3D method is slow, although it would work for a small fraction of the solar volume. 
The computational times for the respective geometries will be compared in section~\ref{sec.comptimes}.

\subsection{Geometrical setup for the $2.5$D problem}

\begin{figure}[b]
\setlength{\unitlength}{1.7mm}
\begin{picture}(40,33)(0,2.5)
\linethickness{0.5mm}
\qbezier(30,15)(30,27)(15,27)
\qbezier(15,7)(30,7)(30,15)
\put(15,7){\line(0,1){20}}
\linethickness{0.2mm}
\put(15,1){\vector(0,1){30}}
\put(10,15){\line(1,0){25}}
\linethickness{0.15mm}
\put(19.5,21){\circle*{0.7}}
\put(15,21){\line(1,0){4.5}}
\put(16,21.6){$\varpi$}
\put(13.5,20.5){$z$}
\put(20.5,20.5){$\tilde{\br}$}
\put(15.9,29){$\zhat$, symmetry axis}
\put(20,10){$\Sigma$}
\put(27.5,7.2){$\partial\Sigma$}
\linethickness{0.1mm}
\put(15,15){\line(3,4){4.5}}
\put(15.5,17.2){$\theta$}
\put(18,17.5){$r$}
\end{picture}
\caption{The thick contour $\partial\Sigma$ delineates the two-dimensional generating section, $\Sigma$, of the volume $V$. A point $\tilde{\br}$ in $\Sigma$ is specified by coordinates  $(r, \theta)$, where $r$ is radius and $\theta$ is co-latitude, or, equivalently,  by coordinates ($\varpi$, $z$), where $\varpi=r \sin\theta$ is the distance to the axis.
}
\label{fig.geometry}
\end{figure}
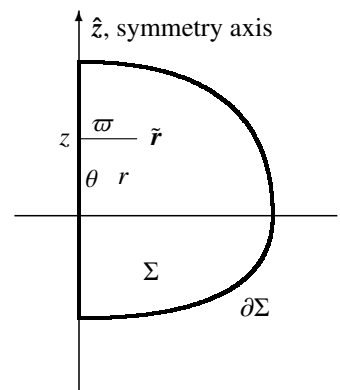

For an axisymmetric background model, the computational domain $\Sigma$ is a meridional generating section of the geometry  (Fig.~\ref{fig.geometry}), i.e. the half-disk of radius $R$ in the case of the sphere.

The background model is symmetric about an axis $\zhat$. In three dimensional space a spatial point is associated with the position vector
\begin{equation}
\br = (r,\theta, \phi)=: (\tilde{\br}, \phi), 
\end{equation}
where $\tilde{\br}$ belongs to the 2D section $\Sigma$ and $\phi$ is longitude about the axis of symmetry (Fig.~\ref{fig.geometry}). The background density, sound speed, and flow are specified at positions $\tilde{\br}$ in $\Sigma$:
\begin{eqnarray}
 \rho &=& \rho(\tilde{\br}) , \\ 
c &=&  c (\tilde{\br}) , \\ 
\bu  &=&  \tilde{\bu}(\tilde{\br}) +    u_\phi(\tilde{\br}) \phihat,
\end{eqnarray}
where $\tilde{\bu}$ is the meridional component of the flow and $u_\phi$ the rotational velocity component.

\subsection{Expansion of solution in longitudinal Fourier modes}

As the unknown of the finite element method we choose the function $\check{\psi}$ defined by
\begin{equation}
\check{\psi} = \rho c \, \psi. 
\end{equation}
The factor $\rho c$ removes the gradients of $\rho$ or $c$ from the  weak form of the equation and thus improves convergence (see Section~\ref{sec.weakform}). The equation solved by $\check{\psi}$ is 
\begin{equation} \label{eq:scalar_psi}
- \frac{\sigma^2}{\rho \, c^2} \, \check{\psi} - \frac{2 \ii \omega}{c} \, \bu \cdot \nabla \left( \frac{\check{\psi}}{\rho c} \right) - \nabla \cdot \left( \frac{1}{\rho} \nabla \check{\psi} \right) = \frac{s}{c}.
\end{equation}
In order to use the $2.5$D solver, the 3D solution $\check{\psi}$ must be expanded as a Fourier series in the longitudinal direction:
\begin{equation}\label{eq:mexpansion}
\check{\psi}(\br,\omega) = \sum_{m=-\infty}^{\infty} \check{\psi}^m(\tilde{\br}, \omega)   \mathrm{e}^{\ii m\phi}.
\end{equation}  
Inserting the above expansion into Eq.~\eqref{eq:scalar_psi}, we see that the three-dimensional problem separates into a set of independent 2D problems, one for each $m$:
\begin{equation}\label{eq.wavem}
- \sigma_m^2 \frac{\check{\psi}^m}{\rho c^2}
 - \frac{2\ii \omega}{c} \tilde{\bu} \cdot \tilde{\nabla} \left( \frac{\check{\psi}^m}{\rho c} \right)
 - \tilde{\nabla} \cdot \left(  \frac{1}{\rho} \tilde{\nabla}  \check{\psi}^m  \right)    =  \frac{s^m}{c} ,
\end{equation}
with
\begin{equation}
\sigma_m^2 = \omega^2  + 2\ii \omega \gamma    - \frac{2 m \omega  \, u_\phi }{\varpi}  - \frac{m^2 c^2}{\varpi^2} 
\end{equation}
and \begin{equation}
s^m (\tilde{\br}, \omega) = \frac{1}{2\pi} \int_0^{2\pi} s(\br,\omega) \, \mathrm{e}^{-\ii m\phi} \diff\phi . 
\end{equation}
The 2D gradient and divergence operators are defined by
\begin{equation}
\begin{aligned}
& \tilde{\nabla} F =  \hat{\varpi} \partial_\varpi F +    \zhat \partial_z F    ,\\
& \tilde{\nabla} \cdot \bF=  \frac{1}{\varpi} \partial_\varpi (\varpi F_\varpi) +     \partial_z F_z .
 \end{aligned} \label{eq.nablah}
\end{equation}
On the outer boundary, we apply a Sommerfeld-like boundary condition
\begin{equation}
\partial_n \check{\psi}^m  = \ii    \frac{\sigma_m}{c} \check{\psi}^m  .
\end{equation}
A boundary condition that would take into account the fast variation of density near the surface (exponential decay)  would be preferable, thus removing the need for an extended atmosphere and resultant mesh; this will be studied in a future paper.

 \subsection{Expansion of Green's function}
 
For a delta function source at position $\br_s = (\tilde{\br}_s, \phi_s)$,
\begin{equation}
s(\br; \bs) = \delta(\phi-\phi_s) \, \delta(\tilde{\br} - \tilde{\br}_s) , 
\end{equation}
and the source coefficients are
\begin{equation}
s^m(\tilde{\br}; {\br}_s) =  \left \{ \begin{array}{l}
\displaystyle \frac{1}{2\pi} \mathrm{e}^{-\ii m \phi_s}  \delta(\tilde{\br} - \tilde{\br}_s) \mbox{ if } \br_s \mbox{ is not on the $\zhat$ axis,} \medskip \\
\displaystyle \frac{1}{2\pi} {\delta_{m,0}} \, \delta(\tilde{\br} - \tilde{\br}_s) \mbox{ otherwise.}
\end{array} \right.
\end{equation}
Thus the Green's function may be computed using
\begin{equation}
 \check{G}(\br,\bs,\omega)  = \sum_{m=-\infty}^{\infty} \check{G}^m(\tilde{\br}, \tilde{\br}_s,\omega) {\rm e}^{\ii m (\phi-\phi_s)} =\rho(\tilde{\br})   c (\tilde{\br}) G(\br,\bs,\omega) ,
\end{equation}
where each $\check{G}^m$ solves 
\begin{equation}
- \sigma_m^2 \frac{\check{G}^m}{\rho c^2}
 - \frac{2\ii \omega}{c} \tilde{\bu} \cdot \tilde{\nabla} \left( \frac{\check{G}^m}{\rho c} \right)
 - \tilde{\nabla} \cdot \left(  \frac{1}{\rho} \tilde{\nabla}  \check{G}^m  \right)    =  \frac{1}{2\pi c} \delta(\tilde{\br}-\tilde{\br}_s).
\end{equation}
If $\tilde{\br}_s$ is on the $z$-axis,    the only non-zero component is  $\check{G}^{m=0}$. Notice that when there is no flow $(\bu= \mathbf{0})$, we have $\sigma^2_{-m}=\sigma^2_m$ and thus $\check{G}^{-m} = \check{G}^{m}$.

\subsection{Weak form of the equations}
\label{sec.weakform}

To implement the finite element method, we derive a weak (variational) formulation of the wave equation for each $m$. Equation~\eqref{eq.wavem} is multiplied by a test  function $\Phi$ and integrated over  $\Sigma$ with integration element $\varpi  \id\Sigma =  \varpi \id\varpi \id z$.

Using integration by parts, we obtain
\begin{eqnarray}\label{eq.weakform}
\begin{aligned}
&
-\int_{\Sigma}     \Phi \,    
\frac{\sigma _m^2}{\rho c^2}
\check{\psi}^m  \,  \varpi \diff\Sigma  
\\& 
- \ii\omega \int_{\Sigma}    
\frac{1}{\rho c^2} \tilde{\bu} \cdot
\left[ (\tilde{\nabla} \check{\psi}^m )\,  \Phi - (\tilde{\nabla} \Phi) \,  \check{\psi}^m
\right] \varpi \diff\Sigma
\\ 
& +  \int_{\Sigma} \frac{1}{\rho} \tilde{\nabla} \Phi \cdot \tilde{\nabla} \check{\psi}^m  \,  \varpi \diff\Sigma
- \oint_{\partial \Sigma}  \frac{1}{\rho} \Phi  \, (\partial_n\check{\psi}^m) \, \varpi \diff  l
\\
 & =   \int_\Sigma     \Phi \frac{s^m}{c}  \,
\varpi \diff\Sigma .
\end{aligned}
\end{eqnarray}
where we have used mass conservation $\tilde{\nabla}\cdot(\rho \tilde{\bu} )  = 0 $ and we have 
assumed that the flow does not cross the computational boundary ($\tilde{\bu} \cdot \hat{\mathbf{n}} =0 \mbox{ on } \partial \Sigma$). The integral $\oint \id l$ is a line integral over the outer boundary $\partial\Sigma$. The boundary condition is used to rewrite the boundary term:
\begin{equation} \oint_{\partial \Sigma}  \frac{1}{\rho} \Phi  \, (\partial_n\check{\psi}^m) \, \varpi \diff l = \ii \oint_{\partial \Sigma}   \Phi  \,\frac{\sigma_m}{\rho  c}\,  \check{\psi}^m \, \varpi \diff l .
\end{equation}

To solve this problem by the FEM, the computational domain is subdivided into quadrilateral cells. On each of them, the test function $\Phi$ and the solution of the equation $\check{\psi}^m$ are projected on a basis $\left\{\Phi_i \right\}$ which, in our study, consists of piecewise continuous polynomials.  By writing this formulation for all the $\Phi_i$ basis functions, the problem becomes a  linear system that can be solved to obtain $\check{\psi}^m$. 
We refer the reader to \citet{FEMbook} for an  introduction to the method, and to \citet{durufle_rr} for more specific details.

\subsection{Comparison with an exact solution for a piecewise homogeneous layered medium}
\label{sec.concShells}
\begin{figure}
\centering
   \includegraphics[width=0.15\textwidth]{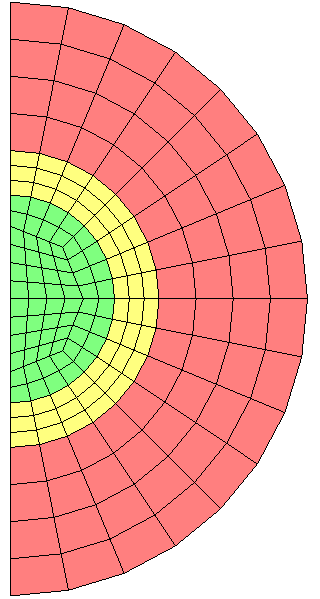}
      \caption{Example of a quadrilateral mesh used to compute the solution for the scattering by spherical layers. 
}         \label{fig.piecewise}
   \end{figure}

In order to test the rate of convergence of the finite-element solver, \cite{durufle_rr} considered a simple benchmark for which the exact solution is known. In this set-up, the background medium consists of a series of concentric spherical shells (with boundaries at radii $r_1 = 0.7 R$, $r_2= R$, and $r_3 = 2 R$, where $R=700$~Mm) where the sound speed $c_i$ and density $\rho_i$ are constant within the $i$-th shell.
An example of a quadrilateral mesh used for the computation can be seen in Fig.~\ref{fig.piecewise}. A Sommerfeld boundary condition is applied at the computational boundary $r_3$. The following coefficients were chosen arbitrarily and are not intended to represent the Sun:
\begin{equation}
\left\{
\begin{aligned}
& \omega  /2\pi = 2 \; \mbox{mHz}   ,   \\
& 1/\rho_1 =  1.5 ,  \qquad 1/\rho_2 =  2  \qquad 1/\rho_3 = 1 , \\
& 1/(\rho_2 c^2_1) =  0.8   ,   \qquad  1/(\rho_2 c^2_2)  =  0.2  ,  \qquad 1/(\rho_3 c^2_3) =  1  ,
\end{aligned}
\right.
\end{equation}
where $\rho_i$ and $c_i$ are in units of $\rho_\odot$ and $700$~km/s respectively.

\begin{figure}
   \centering
   \includegraphics[width=8cm]{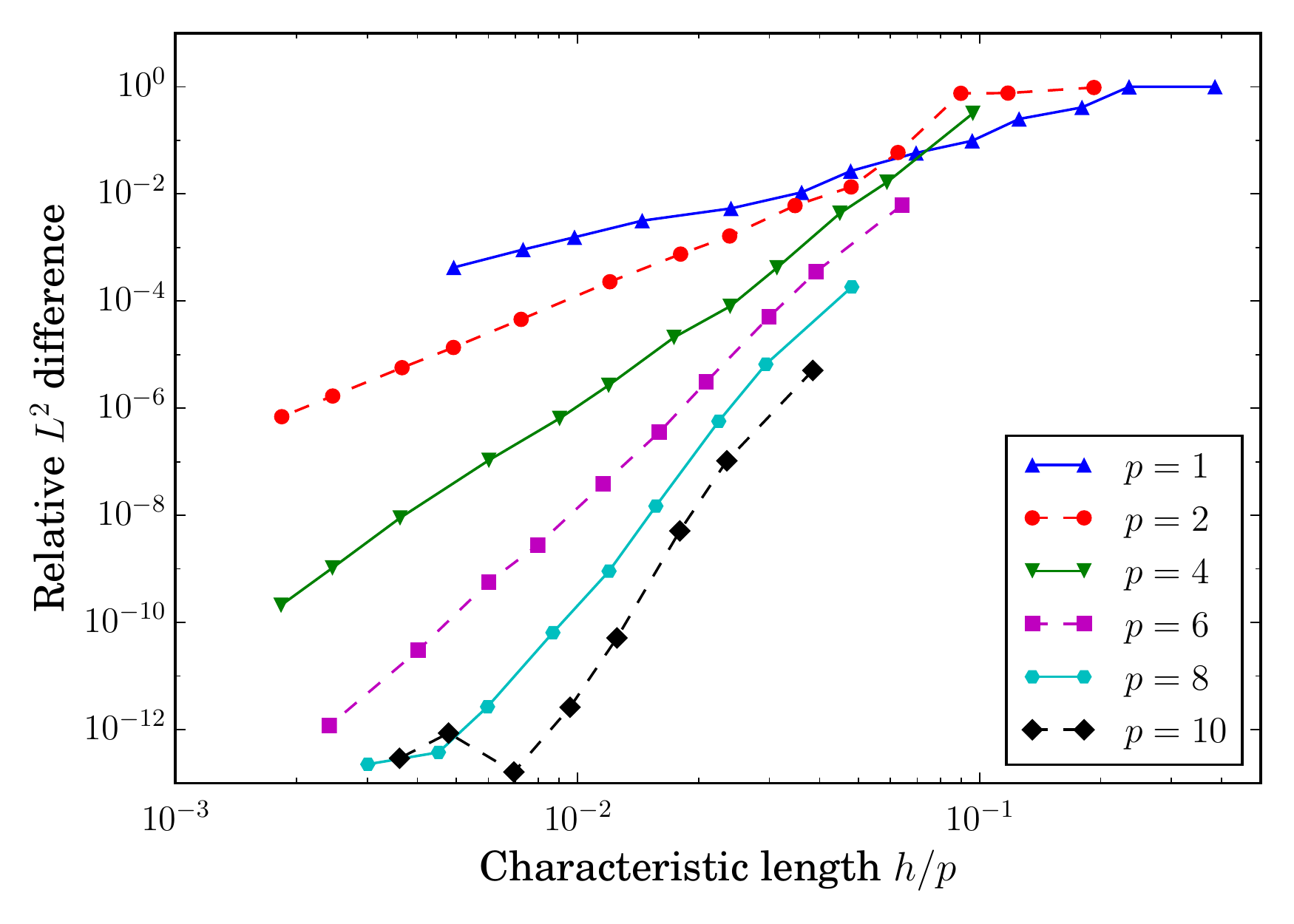}
      \caption{Relative L$^2$ error 
between the FEM solution and the analytical solution for the  piecewise homogeneous layered medium (Sect.~\ref{sec.concShells} and Fig.~\ref{fig.piecewise}) for non-axial incidence of a plane wave at frequency $\omega/2\pi=2$~mHz. The error is plotted as a function of the maximum mesh size $h$ for different 
polynomial orders $p$. The observed rate of convergence is consistent with the theoretical expectation, $h^{(p+1)}$.}
      \label{fig.Fig5}
   \end{figure}

The analytical solution for the scattering of a plane wave traveling from infinity in the +$\xhat$ direction can be computed as a linear combination of spherical Hankel functions \citep{durufle_rr}. 
A non-axial incidence has been selected (the wave vector points in the +$\xhat$ direction) such that all the modes are excited, not only the mode $m = 0$.   
In Fig.~\ref{fig.Fig5}, the relative $L_2$ error between the numerical and the exact solutions is shown.  
The computation of the analytic solution was performed in multiple precision such that the 16 digits of the reference solution are exact.  {We obtain optimal convergence in $h^{p+1}$  where $h$ is the maximum mesh size and $p$ is the order of the polynomial basis \citep{FEMbook}.} The convergence is exponential in $m$ (spectral accuracy). Note that $75$ modes ($|m| \leqslant 37$)  are sufficient to achieve machine precision accuracy for the above problem. 

\section{Time-distance helioseismology in a solar model}
\label{sec.TDH}

\subsection{Spherically symmetric reference model}
\label{sec.smoothModelS}

We use the sound speed and density profiles from the standard solar model described by \citet{jcd_etal_1996}, which is known as Model~S. 
We interpolate the values of  $\rho$ and $c$ on the finite-element grid using  B-splines.

\begin{figure}
\includegraphics[width=\linewidth]{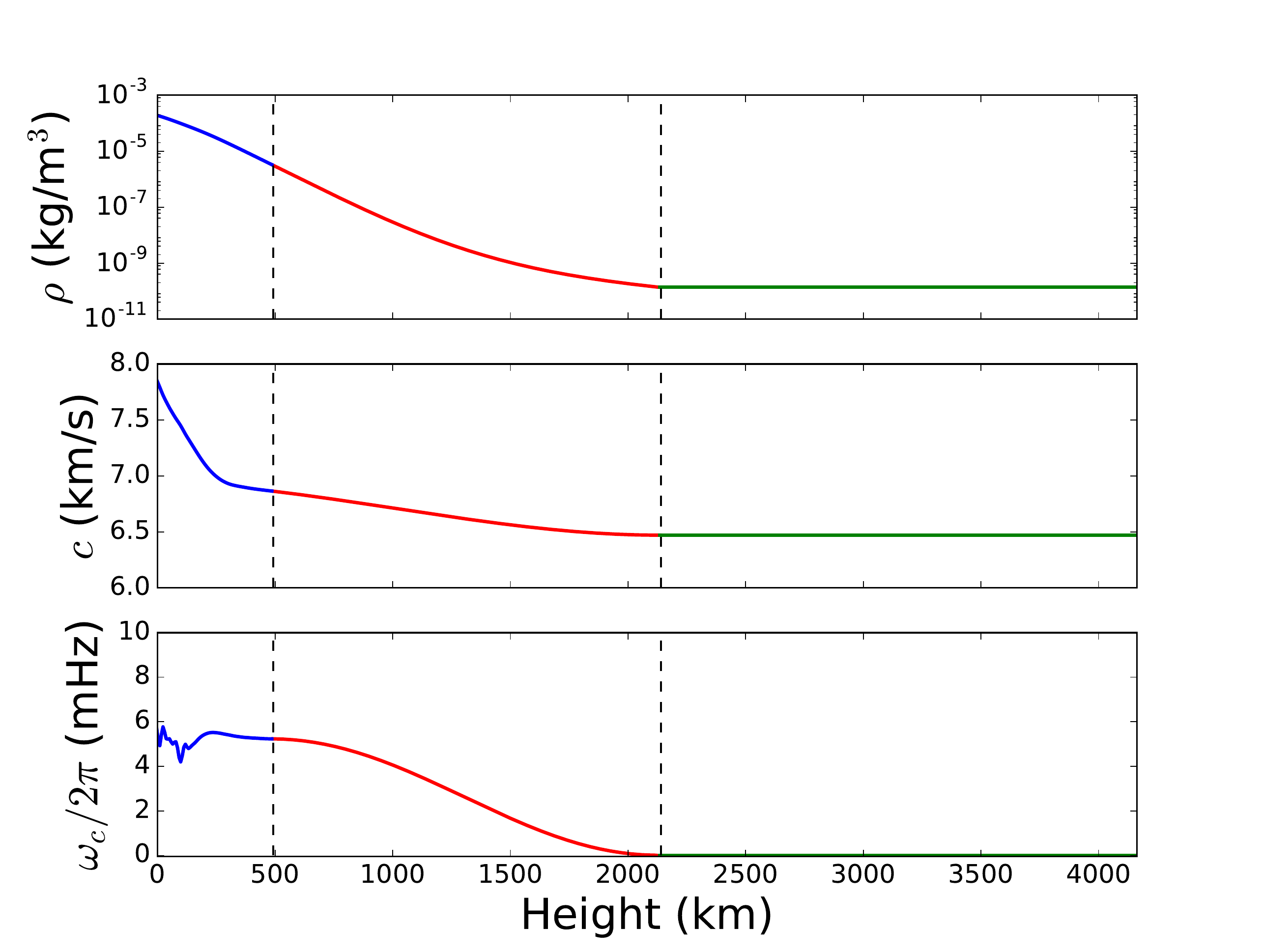}
\caption{The density (top), sound speed (middle) and acoustic cut-off frequency (bottom) of Model S (blue), the transition region (red), and the constant region (green, required for Sommerfeld BC). Height is computed from the photosphere in Model S. The dashed vertical lines indicate the boundaries between regions. }
\label{fig:wac}
\end{figure}

We implement a Sommerfeld-like radiation boundary condition on the computational boundary.
In order to use Eq.~\eqref{eq.sommerfeldBC}, which assumes a locally uniform background medium, we match the Model S atmosphere above $500$~km with a transition region (Fig.~\ref{fig:wac}). For heights between $500$~km and $2100$~km, the acoustic cut-off smoothly transitions to zero and the sound speed to a constant value of $6.5$~km/s, with derivatives vanishing at $2100$~km. The density is deduced by integration using $\omega_c^2 = \rho^{1/2}c^2 \partial_r [ r^2 \partial_r (\rho^{-1/2}) ] / r^2 $ as a definition of the acoustic cut-off frequency.
Above $2100$~km the acoustic cut-off frequency, sound speed, and density are constant.
 By adding such a layer, waves with frequencies above $5.3$~mHz propagate out through this extended region until they are attenuated by the boundary. 
 
To achieve better numerical convergence of the FEM solution, the sound speed and density profiles of Model S are smoothed using B-splines. Eighth-order B-splines are defined by fifteen knots selected to capture all the variations of sound-speed and density throughout the Sun, in particular in the near-surface layers. This ensures that the acoustic cut-off frequency is regular.

\subsection{Meshing the computational domain}\label{sec.mesh}

\begin{figure}
\includegraphics[width=\linewidth]{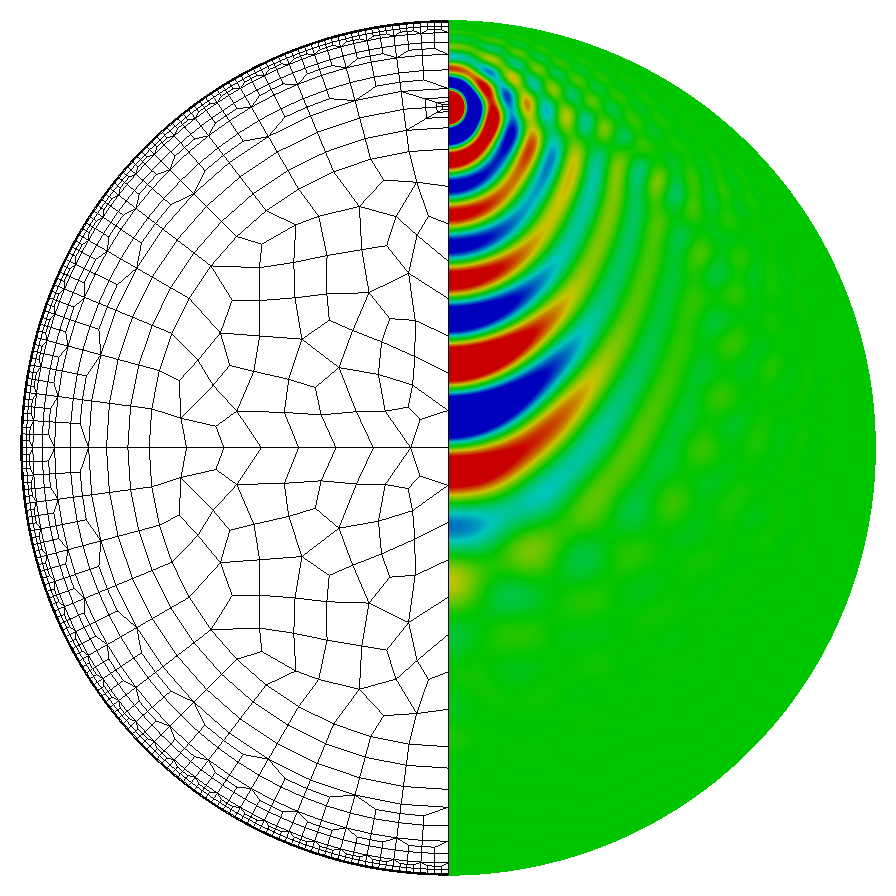}
\caption{Illustrations of (left) the computational mesh used by the finite element method  and (right) the imaginary part of the Green's function at {$\omega/2\pi = 3$~mHz}, with the Dirac source located at radius $0.8R_\odot$ along the $z$-axis. {The damping rate was set to $\gamma/2\pi = 30$~$\mu$Hz.}}
\label{fig.meshgreen}
\end{figure}

\begin{figure}
\includegraphics[width=\linewidth]{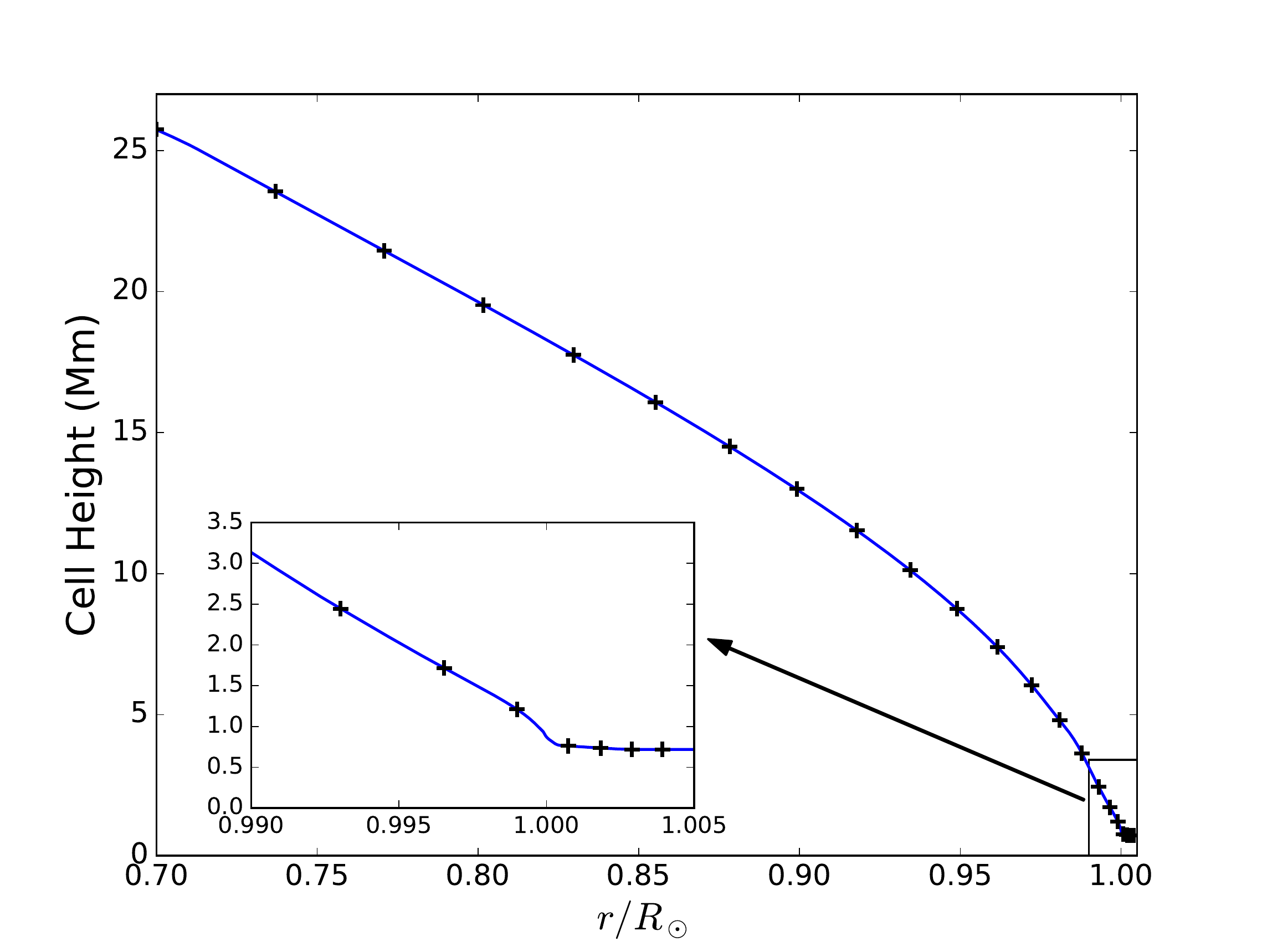}
\caption{The radial wavelength for $\omega/2\pi=9$~mHz and $\ell=15$ (lines) computed from Eq.~\eqref{eq.fullwavelength}, with the position of the cell vertices in the radial direction overplotted (crosses). The local wavelength is captured by a tenth-order polynomial within each cell. Note that the wavelength becomes constant in the extended atmosphere described previously.}
\label{fig.meshsize}
\end{figure}

Figure~\ref{fig.meshgreen} shows the FEM mesh used throughout the remainder of the paper.
Meshing the computational domain is performed in two steps. Initially, we mesh the inner part of the domain ($r<0.7R_\odot$) with quadrilaterals of size $\sim 60$~Mm.
Above this inner mesh we add concentric mesh layers with a radial thickness equal to the radial wavelength 
\begin{equation}\label{eq.fullwavelength}
\lambda_r = 2\pi \left( \frac{\omega^2}{c^2} - \frac{\ell(\ell+1)}{R_\odot^2} \right)^{-1/2} ,
\end{equation}
where $\ell=15$ is the minimum angular degree that we want to study. In the above expression we fix the frequency at $\omega/2\pi=9$~mHz, which is the highest frequency of interest in helioseismology.
The number of points in each mesh cell is determined by the order of the polynomials we choose for the finite elements. In practice, we choose polynomials of order ten in the radial direction, corresponding to a spatial sampling of $\sim \lambda_r /10$. Figure~\ref{fig.meshsize} shows the radial wavelength at $\omega/2\pi=9$~mHz, $\ell = 15$, and selected cell height.
In the horizontal direction, subdivisions are performed such that the horizontal length of the cells is not larger than two times their radial height.
In this work, we use this mesh for all frequencies below $9$~mHz. In future work, one may consider constructing meshes that are less refined for lower frequencies to reduce computational cost.

\subsection{Wave attenuation}
The full width at half maximum (FWHM) of a peak in the observed power spectrum is proportional to  the   attenuation of the mode of oscillation. In our framework the FWHM of a single peak is related to the attenuation through $\gamma = \text{FWHM}/2$, where the FWHM is measured in rad\,s$^{-1}$. Observational studies show that the FWHM of p-mode ridges is both dependent upon the harmonic degree $\ell$ and the frequency \citep[e.g.][]{korzennik_etal_2004}. For this study we restrict ourselves to a frequency dependence only. This approach is acceptable for a filtered power spectrum which selects a wave packet in a narrow range of phase speeds or a single radial order (ridge filtering) because of the  one-to-one relationship between frequency and wavenumber.
However, when modeling the full power spectrum, this approach can only serve as a reasonable estimate. 
While a wavenumber-dependent damping can be implemented in principle, it is beyond the scope of this paper and we reserve this for a future study.

\begin{figure}
\includegraphics[width=0.5\textwidth]{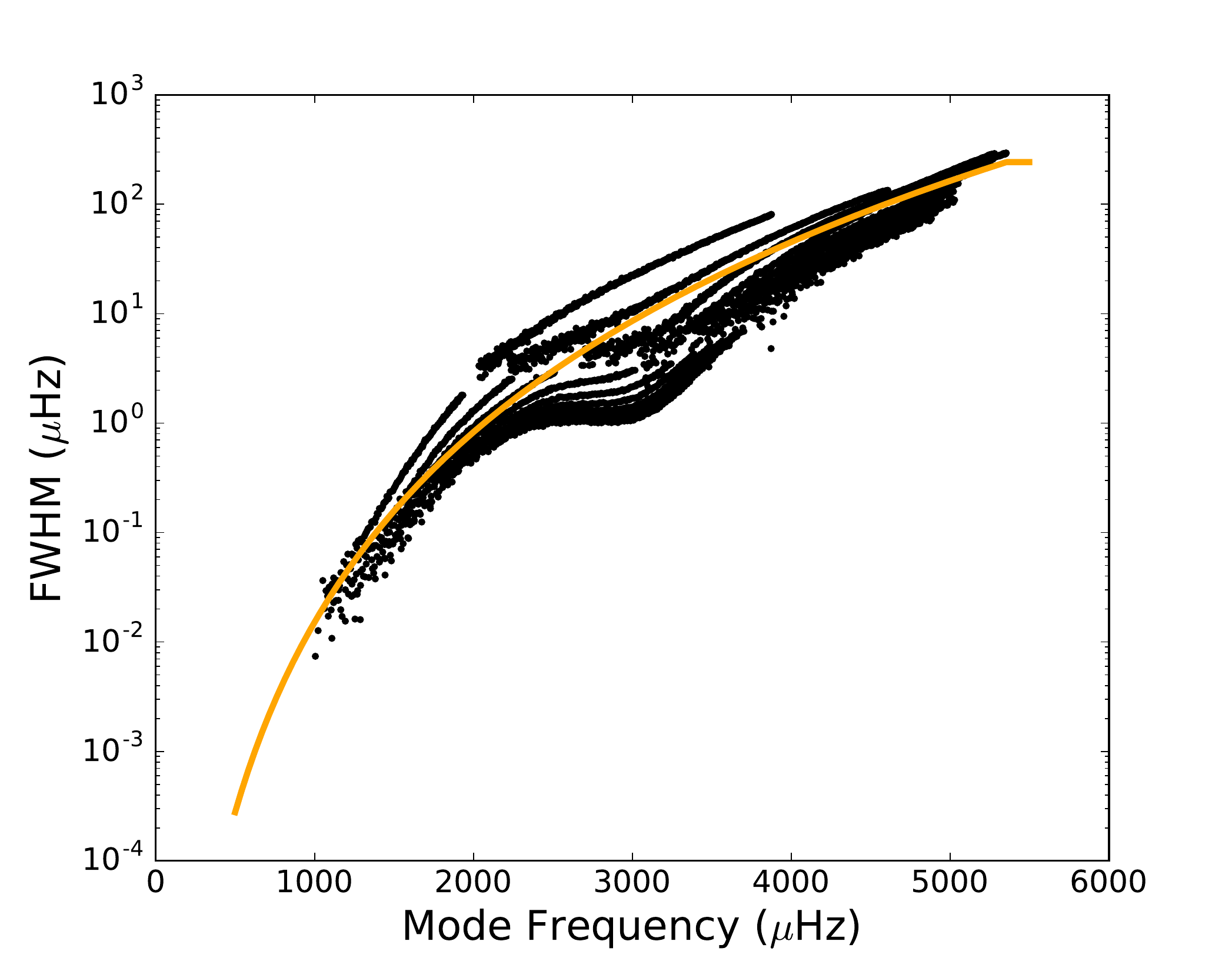}
\caption{The observed full width at half maximum of acoustic modes with radial orders $1\leqslant n \leqslant 22$ \citep[black dots,][]{KORZENNIK_2013, larson_schou_2015},  overplotted with the power law approximation used in the simulations ($2\gamma(\omega)$,   orange line).
}
\label{fig:damping}
\end{figure}

Figure~\ref{fig:damping} shows the observed values of FWHM reported by \citet{KORZENNIK_2013} ($100<\ell<1000$) and \citet{larson_schou_2015} ($\ell<300$) for p modes with radial orders $1\leqslant n \leqslant 22$  and in the range $1$-$5.3$~mHz.
We approximate the attenuation coefficient with a power law in $\omega$ of the form,
\begin{equation}\label{eq.damping}
\gamma(\omega) = 
\gamma_0\left|\frac{\omega}{\omega_0} \right|^\beta  ,
\end{equation}
where $\gamma_0/2\pi = 4.29$~$\mu$Hz, $\omega_0/2\pi= 3$~mHz, and $\beta=5.77$. For frequencies above the acoustic cut-off we fix the attenuation to a constant value of $250$~$\mu$Hz.

\begin{figure*}[t]
\includegraphics[width=\linewidth]{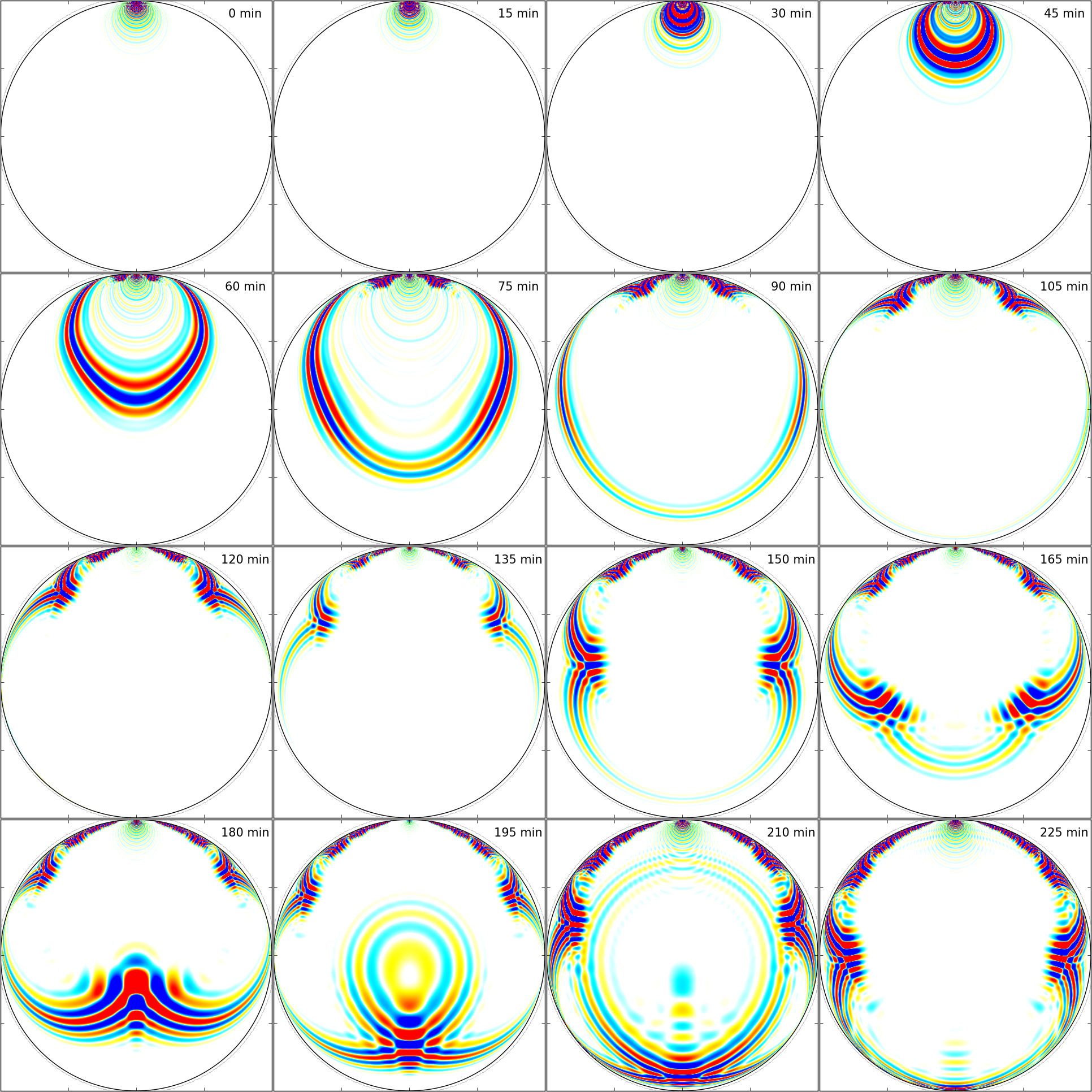}
\caption{Snapshots of  the inverse temporal Fourier transform of $\text{Im} \check{G}(\br, \omega)/\omega$,  propagating from a source located on the polar axis at the photosphere, where $\br$ belongs to a plane through the source and the center of the Sun.
The time $t$, measured from the source time, is written on the top right of each panel.
Each image is multiplied by $\rho^{-1/2} \exp(t/75 \mbox{ min})$ for display purposes. The values are saturated at a hundredth of  the maximum value for the whole time series. The first nine panels have been further saturated (by a factor 2) to improve the visibility of the first arrival wave.}
\label{fig.greens_fcn}
\end{figure*}

\subsection{Green's function}

As the behaviour of the Green's function around a Dirac source is difficult to capture accurately,
the mesh is refined around the position of the source, which, for numerical stability of our solver,
has to coincide with an existing mesh point.
This refinement is performed by subdividing the cells neighboring the source point into three quadrilaterals
such as the shape of the cells around the source is preserved (see Fig.~\ref{fig.meshgreen}).
In practice, we subdivide the cells around the Dirac source five times.

With all the tools in hand, the Green's function can be computed for any given frequency. Figure~\ref{fig.greens_fcn} shows snapshots in the time domain  of the inverse Fourier transform of $\text{Im} G/\omega$, for a source located on the $z$-axis at the Solar surface. For this figure we computed $5000$ equidistant frequencies (from 0 to 8.33~mHz)
in order to cover a time span of about $7$ days. In the first six panels the first-arrival wave front is seen propagating away from the source through the core towards the far side of the Sun.  At $t=75$~min, the second-skip waves become visible. Wave packets with higher skip numbers (which take longer to travel) are also seen at the latter times. The first-arrival wave packets reaches the farside of the sun around 135~min and is seen at time $t=165$~min propagating back towards the source.

\subsection{Oscillation power spectrum}
In Sect.~\ref{sec.convSource} we demonstrated that under certain assumptions the cross-covariance can be directly related to the Green's function. But does such a source of excitation produce reasonable oscillation power spectra? This is an important test as the cross-covariance function is directly connected to the oscillation power spectrum \citep[e.g.][]{Sekii2003}. Here we compute  the power spectrum in terms of Im~$G$ and compare with observations, in the case of a spherically symmetric background medium.

Consider the observable measured at radius $R$ and take its  spherical harmonics transform:
\begin{equation}
\psi_\ell^m(\omega) = \oint \id\rhat_0 \, Y_\ell^{m*}(\rhat_0) \int_V   G(R\rhat_0, \br, \omega) \, s(\br, \omega)   \, \rho \id\br  ,
 \end{equation}
 where $\diff\rhat_0 = \sin\theta_0 \diff\theta_0 \diff\phi_0$ is the surface element on the unit sphere.
The power spectrum is then defined by the expectation value of the squared modulus of the observable,
\begin{equation}\label{eq.projCSphHrm}
\begin{aligned}
\mathscr{P}_\ell^m(\omega)
&= \EE[ |\psi_\ell^m(\omega)|^2] \\
&=  \oint \diff\rhat_0 Y_\ell ^{m}(\rhat_0)  \oint \diff\rhat'_0  Y_\ell^{m*}(\rhat_0')  \; \overline{C}(R\rhat_0 , R\rhat'_0, \omega)       .
\end{aligned}
\end{equation}
We rewrite $Y_\ell^m(\rhat_0')$  in a frame ${\cal R}_0$ 
with polar axis  $\rhat_0=(\theta_0,\phi_0)$. In this frame, we denote by $\rhat=(\Theta, \Phi)$ the polar angles of $\rhat_0'$. The rotation 
of Euler angles $(\alpha,\beta,\gamma)=(\pi,\theta_0,\pi-\phi_0)$ brings  ${\cal R}_0$ 
to the original frame. Using the rotation formula of spherical harmonics \citep[e.g.,][]{Messiah1960}, we have
\begin{equation}
\begin{aligned}
& \mathscr{P}_\ell^m (\omega)= \\
&=   \oint \id\rhat_0 Y_\ell^{m}(\rhat_0)  \oint \id\rhat  \left(  \sum_{m'=-\ell}^\ell  Y_\ell^{m'}(\rhat)  {\cal D}_{m' m}^{(\ell)}(\alpha, \beta, \gamma)    \right)^*  
\overline{C}(R \zhat, R\rhat,  \omega) 
 \\
&=  \left( \oint \diff\rhat_0   \,  Y_\ell^{m}(\rhat_0)   {\cal D}_{0m}^{(\ell)*}(\alpha, \beta, \gamma)  \right)  
  \oint \diff\rhat \,Y_\ell^0(\rhat)   \, \overline{C}(R\zhat, R\rhat, \omega)    ,
\end{aligned}
\end{equation}
where, for the sake of simplicity,  $\overline{C}$ was assumed to depend only on  angular distance $\Theta$  (horizontal isotropy and $\bu=0$). 

\begin{figure}[htb]
\centering
\subfloat{\includegraphics[width=\linewidth,trim={0 0 1cm 0},clip]{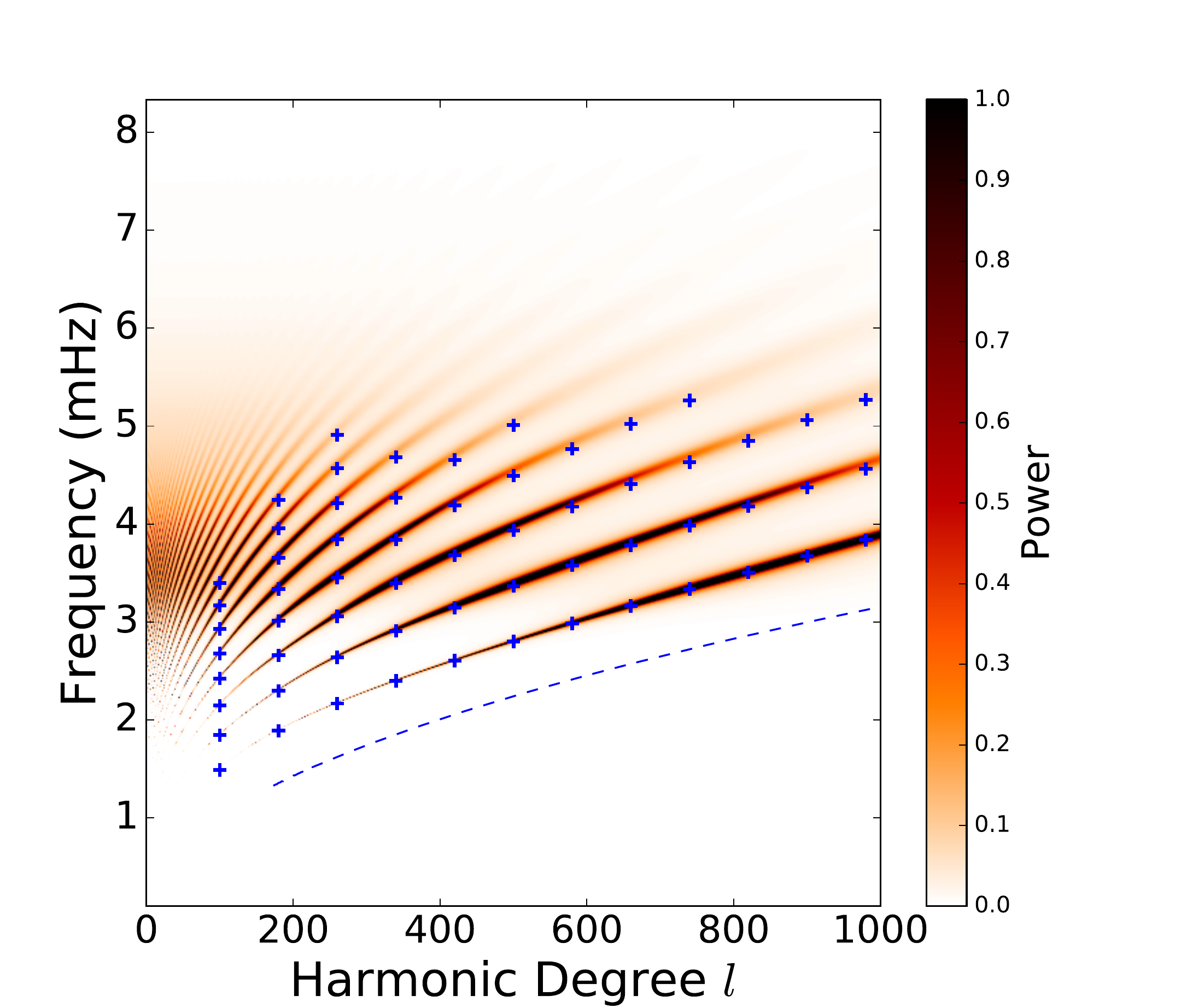}}\\
\subfloat{\includegraphics[width=\linewidth,trim={2cm 0 1cm 0},clip]{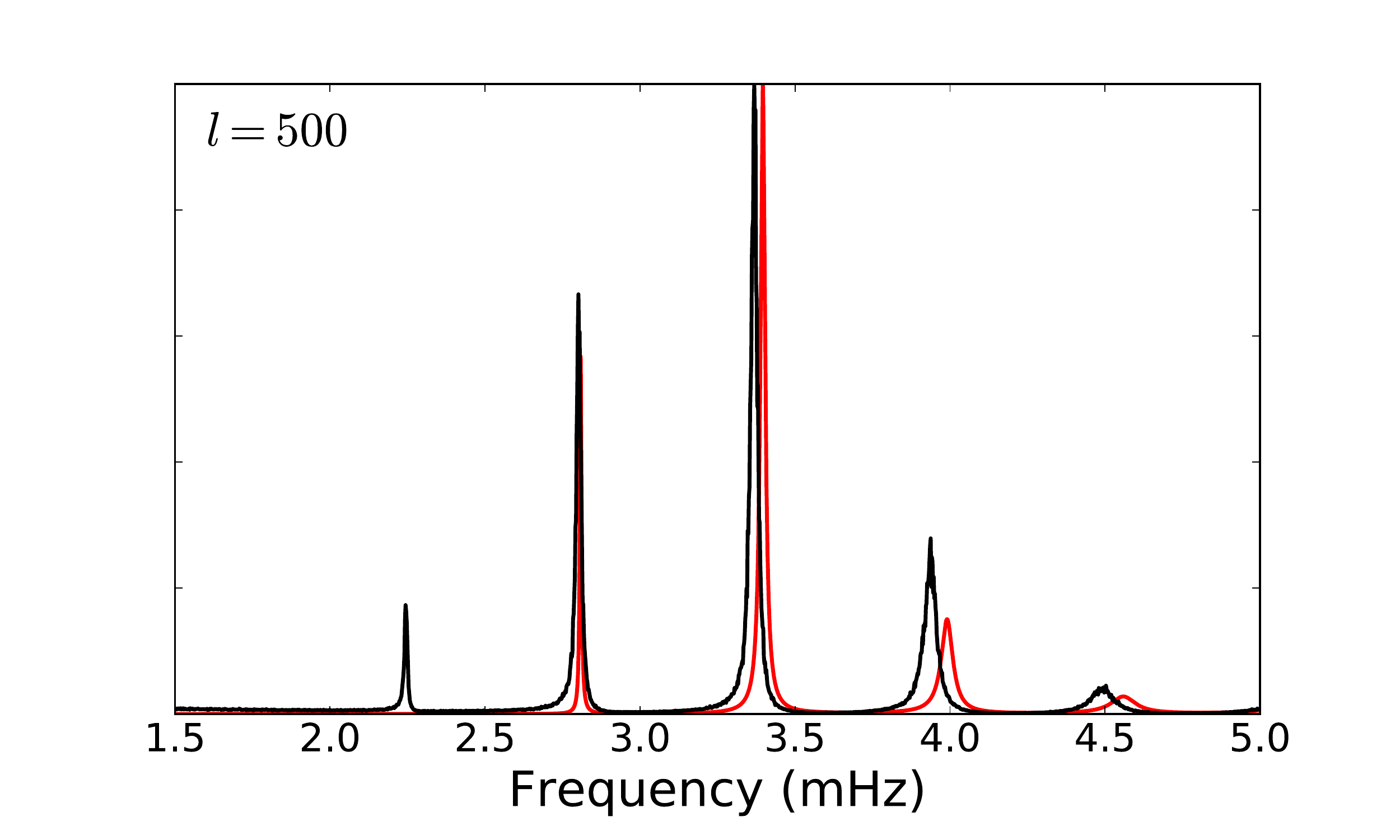}}
\caption{Top: Power spectrum  computed at the solar surface. The blue crosses indicate the position of the p$_1$-p$_8$ modes reported by \citet{KORZENNIK_2013}, while the blue dashed line shows the observed f-mode ridge which is missing in this simulation due to the lack of a gravitational term. Bottom: Comparisons of the simulated power spectrum (red) and HMI data (black) at $\ell = 500$. The small misalignment of the simulated ridges from the observations are due to imperfect modeling of surface layers in Model S \citep[i.e., surface effects,][]{rosenthal_etal_1999}.}
\label{fig.powerspect}
\end{figure}

Using the explicit form of the rotation matrix elements, 
\begin{equation}
{\cal D}_{0m}^{(\ell)}(\alpha, \beta, \gamma) 
= (-)^m \sqrt{\frac{4\pi}{2\ell+1}} Y_\ell^{m*}(\beta,\gamma)
= \sqrt{\frac{4\pi}{2\ell+1}} Y_\ell^{m}(\rhat_0) ,
\end{equation}
 the expression for the power spectrum simplifies to
\begin{eqnarray}\label{eq.power}
\mathscr{P}_\ell^m(\omega)
&=&   \sqrt{\frac{4\pi}{2\ell +1}}  \oint \id\rhat \,Y_\ell^0(\rhat)   \, \overline{C}(R \rhat, R \zhat, \omega) \nonumber  \\
&=&   2\pi  \int_0^\pi P_\ell(\cos\Theta)   \overline{C}(\Theta, \omega) \sin\Theta \diff\Theta
\nonumber  \\
&=&    \frac{\Ps(\omega)}{\omega}
  \int_0^\pi  {\rm Im} G(\Theta, \omega)    \, P_\ell(\cos\Theta)   \,\sin\Theta \diff\Theta , 
  \end{eqnarray}
 where the last equality is for the special source without flow. The function $P_\ell$ is the Legendre polynomial of order $\ell$.

We have complete freedom in the choice of the frequency dependence of the source power, $\Ps(\omega)$.
In the rest of the paper we choose a Lorentzian profile:
\begin{equation}
\Ps(\omega) = \left[1+\left(\frac{|\omega| - \omega_0}{\Gamma/2}\right)^2\right]^{-1} ,
\end{equation}
where $\omega_0/2\pi=3.3$~mHz and $\Gamma/2\pi = 1.2$~mHz. This choice is reasonable for the purposes of this paper.

For a source on the polar axis, only the $m=0$ mode needs to be computed.
To avoid aliasing, the Green's function is sampled on a high-resolution grid in $\theta$ to increase the spatial  Nyquist frequency.  In these results, we used 20000 grid points in $\theta$.

\begin{table*}
  \centering
  \caption{Values of the $a$-coefficients  for the $\ell=85$ and $n=8$  mode near  $3.2$~mHz computed from Montjoie simulations, ADIPLS eigenvalue calculations, and measured from SDO/HMI observations. The frequency resolution of the Montjoie simulations is $\Delta\nu=0.5$~$\mu$Hz. Simulation \#1 was performed using Eq.~\eqref{eq.scalar} and simulation \#2 includes the second-order term $\bu\cdot\nabla(\bu\cdot\nabla\psi)$. For the ADIPLS calculations, the odd $a$-coefficients are obtained from the rotational sensitivity kernels. The modified ADIPLS calculation ignores gravity terms. See main text.}
  \label{fig.mpowerspectTable}
  \begin{tabular}{rrrrrr}
   \hline\midrule      
   $a$-coefficients & Simulation~\#1 & Simulation~\#2 & Modified ADIPLS & ADIPLS & SDO/HMI \\
     $(\ell,n)=(85,8)$ 
    & (nHz) & (nHz) & (nHz) & (nHz) & (nHz) \\\midrule
   $\overline{\nu}_{ln}$ &  $ 3\,218\,437.9$ & $3\,218\,150.0$ &$3\,221\,934.5 $& $3\, 215\, 796.6$ & $ 3\,205\,271.5 \pm 7.2 $ \\
    $a_1$  & $437.8$ &$437.0$&$436.8$& $435.8$ & $442.9 \pm 0.1$ \\
    $a_2$  & $1.8$   &$0.0$  &$0.0$  & $0.0$ & $0.6   \pm 0.2$\\
    $a_3$  & $17.9$  &$19.0$ &$20.7$ & $20.7$ & $22.1  \pm 0.2$ \\\midrule
  \end{tabular}\normalsize
\end{table*}

Figure~\ref{fig.powerspect} shows the $m=0$ power spectrum $\mathscr{P}_\ell^0(\omega)$ computed from Eq.~\eqref{eq.power} with the source located at the photosphere on the $z$-axis. Here we have computed $8000$~frequencies from $0$ to $8.3$~mHz and harmonic degrees up to $1000$.  This figure shows a good relationship between the mode frequencies of our simulation and those of MDI/Doppler measured by \citet{KORZENNIK_2013}. Unlike the normal-mode summation method used in previous work, our power spectrum shows physical ridges above the cut-off frequency. Figure~\ref{fig.powerspect} also shows a slice through the power spectrum at $\ell=500$ compared with 72-days of observations from HMI/SDO.  We note that at high frequencies the mode frequencies are slightly larger than the
observed values. This is due to imperfect modeling of the surface layers in Model S (surface effects), not to numerical issues (the accuracy of the Green's function is discussed later).

\subsection{Frequency splittings due to differential rotation}
Having demonstrated the agreement between our simulations and observations in the case of no background flow, we now turn our attention to solar rotation. We wish to check that the differential rotation of the Sun's convection zone will introduce the correct frequency splittings  between the azimuthal modes propagating in the prograde ($m>0$) and retrograde ($m<0$) directions. 
We compute the Green's function for a photospheric source $\one$ located at the equator at longitude $0^\circ$, in the presence of a flow $\bu = \Omega(r,\theta)\varpi\phihat$.
We use a solar-like differential rotation model specified by 
\begin{equation}
  \frac{\Omega(r,\theta)}{2\pi} = 
  \left\{ 
  \begin{array}{ll}
  454 - 55 \cos^2\theta - 76 \cos^4\theta
  \text{  nHz } &  r>0.7R_\odot , \\
 435\text{  nHz } &  r< 0.7 R_\odot. 
 \end{array}
 \right.
 \label{eq.rotprofile}
\end{equation}

\begin{figure}
  \centering
  \includegraphics[width=\linewidth]{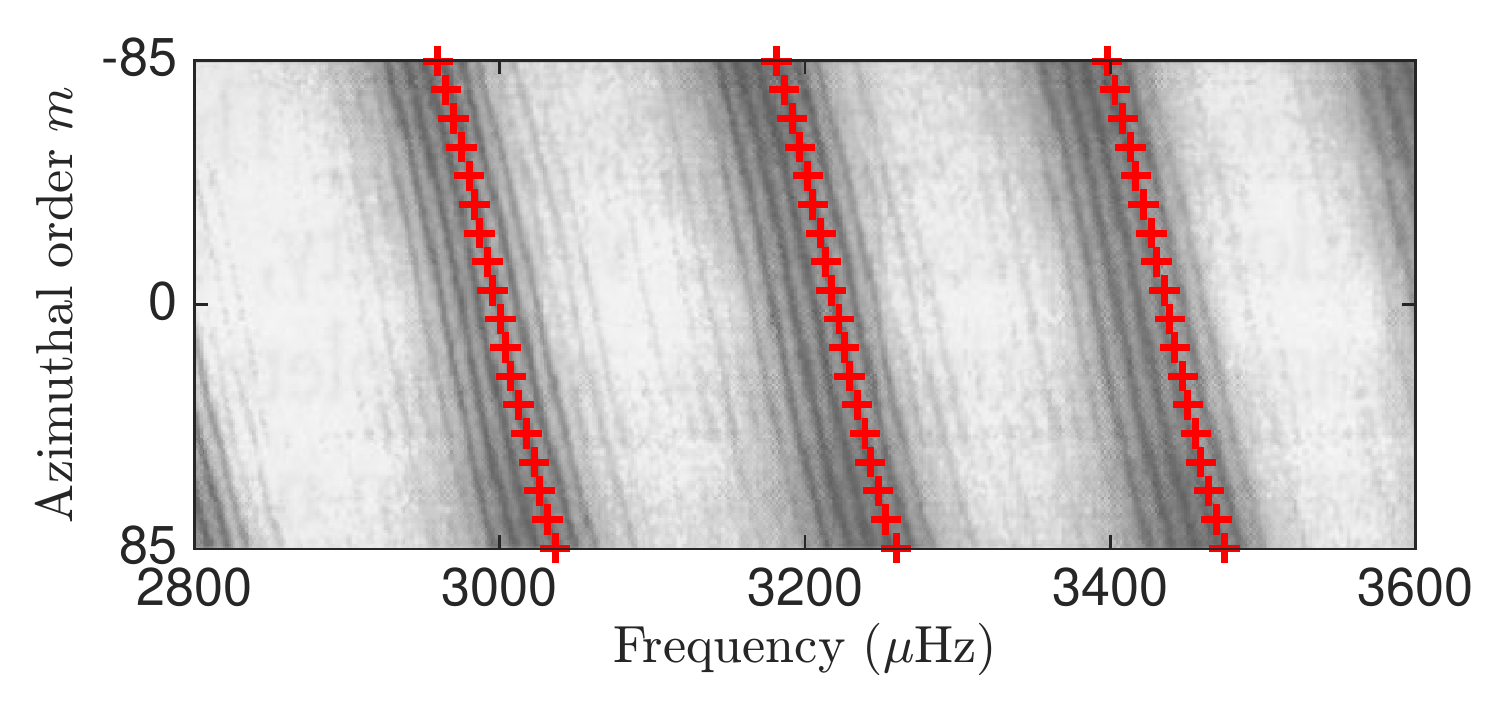}
  \caption{Section of the power spectrum $\mathscr{P}_\ell^m(\omega)$ for harmonic degree $\ell=85$ around frequency $3.2$~mHz ($n=7,8,9$). Red crosses are mode frequencies of the rotationally split p-modes from the first Montjoie simulation in Table 1. 
  {For comparison, the gray scale image shows the GONG  observational power spectrum taken from the paper by \citet{Hill1996}. Darker shades indicate larger values of the power.} Note the side lobes in the observations due to aliasing from observing half of the Sun.   The slope ($a_1$) of the frequencies with $m$ is due to the average rotation rate, while the curvature  ($a_3$) indicates differential rotation (slower rotation near the poles).}
  \label{fig.mpowerspect}
\end{figure}

For comparison with the $\ell=85$ GONG power spectrum near $3.2$~mHz as reported by \citet{Hill1996}, we compute the Green's function for frequencies between $2.8$ and $3.6$~mHz in steps of 0.5~$\mu$Hz for all azimuthal orders $|m| \leq \ell$. For each value of $m$ and $\omega$, a power spectrum is computed by projecting the cross-covariance $\overline{C}(\br,\omega)$ onto spherical harmonics as in Eq.~\eqref{eq.projCSphHrm}.
The frequencies of the modes with radial orders $n=7,8,9$ were then extracted from each $\mathscr{P}_\ell^m(\omega)$ by fitting  Lorentzian functions.
These mode frequencies are plotted in Fig.~\ref{fig.mpowerspect} over the  observational GONG power spectrum from \citet{Hill1996}.

In order to  quantitatively characterize the frequency splittings due to rotation, we compute the $a$-coefficients as defined by~\citet{Schou1994}. The mean frequency of the multiplet  $\ell=85$ and $n=8$ and the first three $a$-coefficients are given in Table~\ref{fig.mpowerspectTable} in five cases: 
\begin{enumerate}
\item
Montjoie simulation using scalar wave equation~\eqref{eq.scalar} with a surface delta-function source at the equator. The $a$-coefficients are extracted from fits to the mode frequencies measured from the simulated power spectrum.
 \item
 Montjoie simulation including the second-order term $\bu\cdot\nabla(\bu\cdot\nabla\psi)$ on the left-hand side of Eq.~\eqref{eq.scalar}. We observe that the $a_2$ coefficient (asphericity) vanishes.
 \item
Eigenvalue calculation for Eq.~\eqref{eq.scalar} with (non-rotating) Model S and a free surface boundary condition at height $0.0007 R_\odot$ above the photosphere, using a modified version of ADIPLS \citep{Christensen-Dalsgaard2008} to compute the eigenfrequencies $\overline{\nu}_{\ell n}$ and the rotational kernels \citep{JCD}. The modifications to ADIPLS are explained in Appendix~\ref{app:ADIPLS}. The odd $a$-coefficients are derived from the first-order perturbation to the mode frequencies. The even $a$-coefficients are zero to this level of approximation.
\item 
Eigenvalue calculation using the standard ADIPLS pulsation code, without neglecting terms in the oscillation equations.
 \item
Measurements of $a$-coefficients from 360 days of SDO/HMI observations~\citep{larson_schou_2015}.  
\end{enumerate}

\begin{figure*}[t]
\centering
\includegraphics[width=0.8\linewidth]{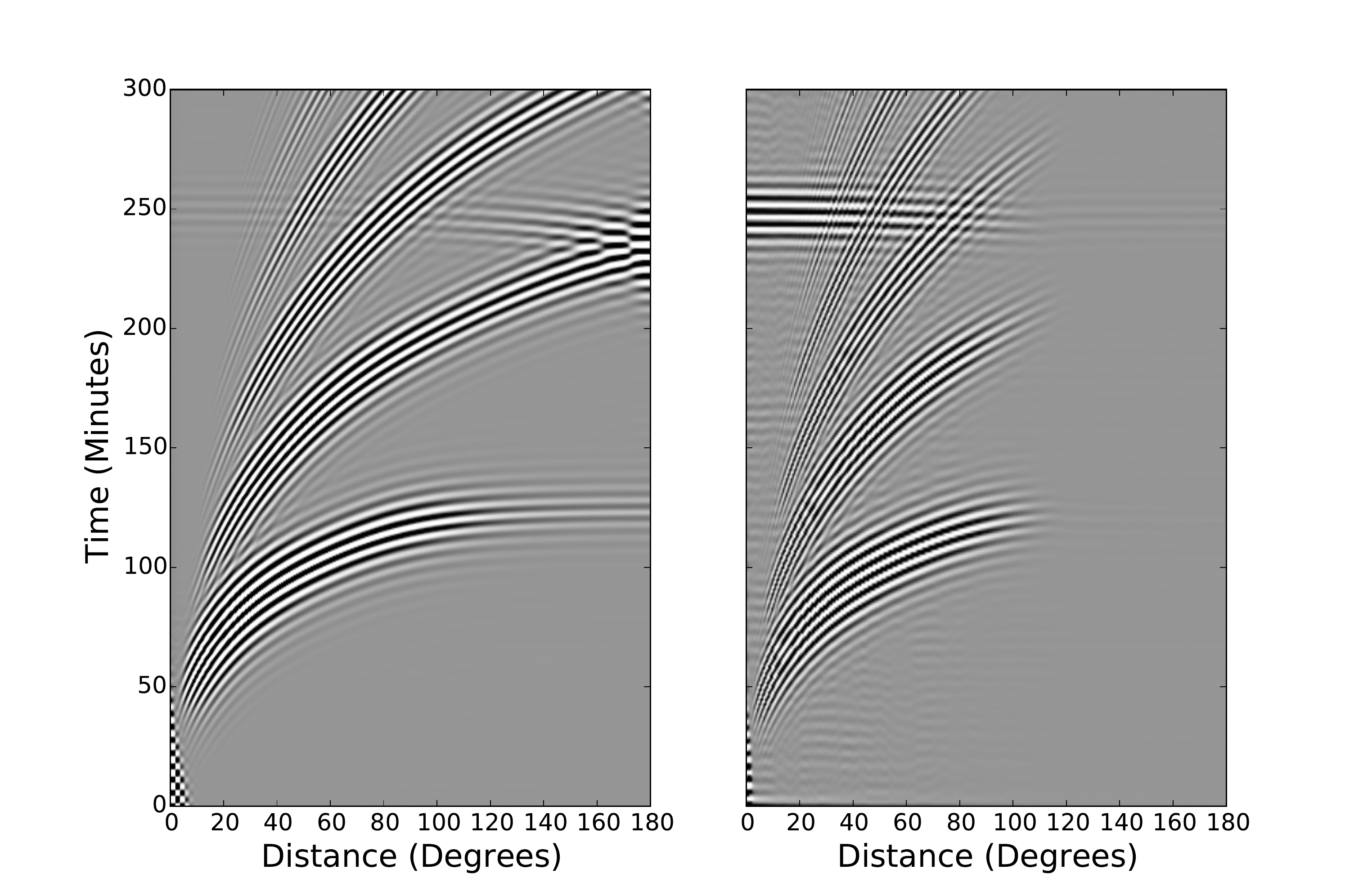}
\caption{Left: Time-distance diagram computed from Eq.~\eqref{eq:TD} at the height of the Dirac source. The right panel is the observational time-distance diagram computed from the Fourier transform of the SOHO/MDI/Doppler medium-degree power spectrum \citep{kosovichev_etal_1997}. The SOHO/MDI time-distance diagram fades away at large separation distances due to foreshortening.}
\label{fig.TD}
\end{figure*}

As mentioned previously, the mean frequencies of Model~S using Montjoie overestimate those of the SDO/HMI observations by $\sim13~\mu$Hz. The ADIPLS mean frequencies are also above the SDO/HMI observations by more than $10$~$\mu$Hz. This difference comes from imperfect modeling in the near-surface layers and is often referred to as 'the surface effect' \citep[e.g.][]{rosenthal_etal_1999}. 
{The $\sim3$~$\mu$Hz frequency difference between the Montjoie and the modified ADIPLS calculations comes from the difference in the atmospheric models. } The difference between the modified ADIPLS and the standard ADIPLS frequencies comes from neglecting the buoyancy force in Eq.~(\ref{eq.Hmaster}).

The simulated $a_1$ and $a_3$ coefficients are of the expected sign and order of magnitude, within a few nHz of each other. The simulated $a_1$ coefficients are about $5$~nHz smaller than the SDO/HMI observed value, even though we did not tune the rotation profile in the simulations. The simulated $a_3$ coefficients are also smaller than the observed value, by $\sim 2$~nHz. 

The value of $a_2$ in  simulation \#1  using Eq.~\eqref{eq.scalar} is non-zero, which was (at first) unexpected since our model does not include centrifugal distortion. This is due to the fact that eigenfunctions are affected by rotation at first order and thus leaves a signature in the power spectrum. Adding the  term $\bu\cdot\nabla(\bu\cdot\nabla \psi)$ in simulation \#2 restores the east-west antisymmetry of the advection of the waves by the flow. 

Overall, this comparison between simulated and observed mode frequencies is very encouraging.

\subsection{Time-distance diagram}
For a spherically symmetric solar model, the expectation value of the cross-covariance function in the time domain is
\begin{eqnarray}
\label{eq:TD}
\overline{C}(\Theta, t) & = & \int_{-\infty}^\infty \overline{C}(\Theta, \omega)  \, {\rm e}^{-\ii \omega t} \,\id\omega \nonumber \\
 &= & \int_{-\infty}^\infty   \frac{\Ps(\omega)}{2\omega}  \, \text{Im} G(\Theta, \omega)\, {\rm e}^{-\ii \omega t}  \, \id\omega, 
\end{eqnarray}
where $\Theta$ is the angular distance on the surface between the two observation points. 
The cross-covariance function is also called the time-distance diagram after  \citet{Duvall1993}. In Fig.~\ref{fig.TD} we compare the time-distance diagram computed from our power spectrum to an observed time-distance diagram using SOHO/MDI medium degree data \citep{kosovichev_etal_2000}. In order to make this comparison we applied a spatial filter to the simulated power spectrum, $F_\ell = [1-\tanh\left(0.03 \ell-3\right)]/2$ for $\ell < 100$ and 0 otherwise, to remove high-degree modes.

\begin{figure*}[hbt]
\centering
\includegraphics[width=0.8\linewidth]{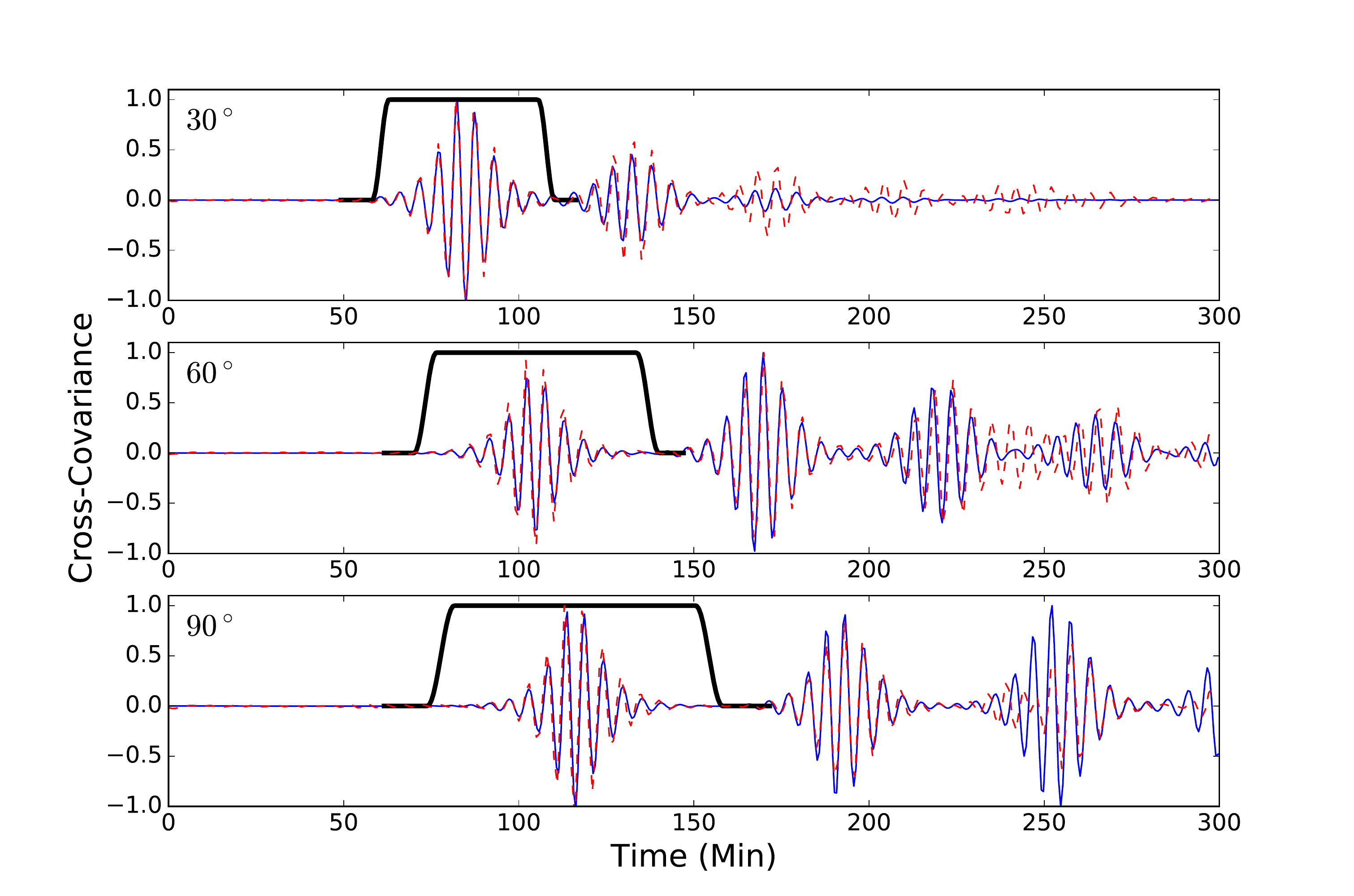}
\caption{Temporal cross-covariance function for three angular distances $\Theta=30^\circ$, $60^\circ$, and $90^\circ$ for the simulations (blue) and the SOHO/MDI Doppler observations \citep[red dashes,][]{kosovichev_etal_2000}. The temporal window functions ($w$) used  in the definitions of travel times are shown in black.}
\label{fig.TDslices}
\centering
\includegraphics[width=0.8\linewidth]{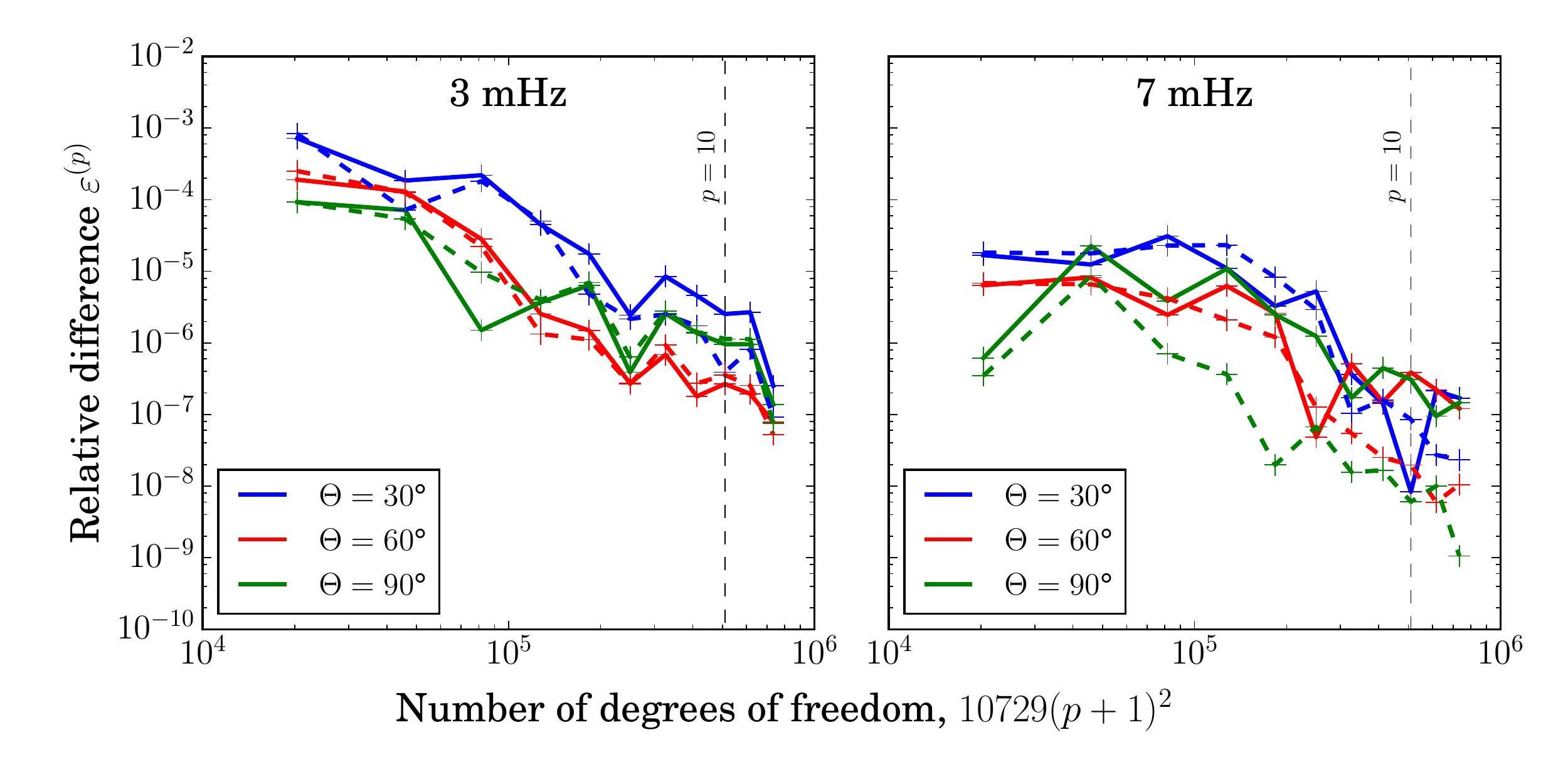} 
\caption{Relative difference between $\mathrm{Im}\, G^{(p)}(\Theta,\omega)$ and $\mathrm{Im}\, G_{\rm ref}(\Theta, \omega)$ as defined by Eq.~\eqref{eq:defDiffXS} for frequencies $\omega/2\pi=3$~mHz (left) and $\omega/2\pi=7$~mHz (right), and different angular distances $\Theta=30^\circ$, $60^\circ$ and $90^\circ$. Computations were performed without flow (solid lines) and with a meridional flow (dashed lines).} \label{fig:diffXS} 
\end{figure*}

Comparisons of the two time-distance diagrams is encouraging. However, the amplitude of the back-skip ridge at $t\sim 250$~min
is greater in the observations than in the simulations, for which we have no definitive explanation. We think that the most likely explanation is that the damping of the low degree modes is  overestimated in our model, resulting in a reduced amplitude of the back-skip branch in the time-distance diagram. Further tuning of the power spectrum is required in order to resolve this discrepancy. To further compare with the observations, Fig.~\ref{fig.TDslices} shows time plots at three different travel distances. The widths and relative amplitudes of the first few skips are in general agreement with the observations.

\section{Validation for helioseismology applications}

\subsection{Convergence of Green's function} 
\label{sect.convergenceGreen}
\renewcommand{\Im}{\mathrm{Im}\,}
\newcommand{\GREF}{G_\mathrm{ref}}
\newcommand{\CREF}{C_\mathrm{ref}}
\newcommand{\ErrorG}{\varepsilon^{(p)}}

In order to estimate the accuracy of the forward solver, we first measure  the convergence of the Green's function towards a solution $\GREF$ computed for a highly refined mesh (four times more cells) with high-order discretization (order 13). This solution is used as reference since we cannot determine the exact solution to our problem. By choosing a basis of polynomials of order 13 for the finite elements on the refined mesh, the number of degrees of freedom per wavelength is 26 ($13\times 2$ cells per wavelength), i.e. many more than the 10 points per wavelength used previously.

We compute the Green's function $G^{(p)}$ by choosing polynomials of order $p$ in the non-refined mesh containing $10729$ cells. The relative difference between $\Im G^{(p)}$ and $\Im \GREF$, denoted by $\ErrorG$,  is plotted in Fig.~\ref{fig:diffXS} as a function of the number of degrees of freedom $N^{(p)}=10729 (p+1)^2$. Explicitly,
\begin{equation}\label{eq:defDiffXS}
  \ErrorG(\Theta,\omega)  = \frac{| \Im G^{(p)}(\Theta,\omega) -\Im G_{\rm ref}(\Theta,\omega)  |}  {\|  \Im G_{\rm ref} (\cdot ,\omega) \|_{L_2} }.
\end{equation}
In Fig.~\ref{fig:diffXS}, we plot $\ErrorG$ for different angular distances between the source and the receiver $\Theta$ ($30^\circ$, $60^\circ$, and $90^\circ$) and different frequencies ($3$~mHz and $7$~mHz). We see that $\ErrorG$ reaches $\sim 10^{-5}$ for orders of discretization $p>10$. We obtain a similar convergence when a meridional flow, as described in Appendix~\ref{app:flowCell} (with  surface velocity $U=20$~ms$^{{-1}}$), is added to the background.


\begin{figure}[htb]  
\includegraphics[width=\linewidth] {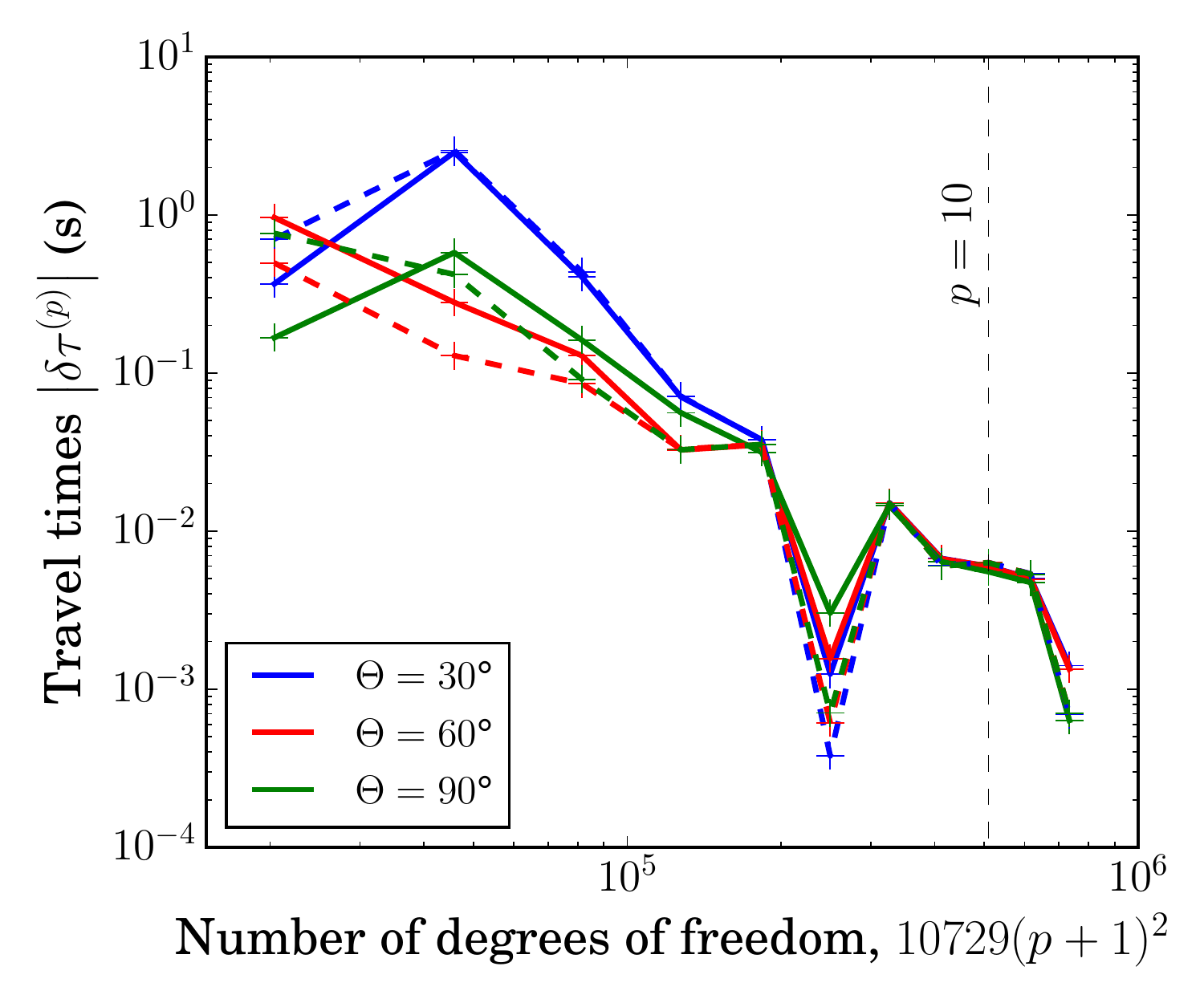} 
\caption{Travel times $\delta\tau^{(p)}$ as defined in Eq.~\eqref{eq:def_tt_cicref}, computed from the difference between cross-covariances $C^{(p)}(\Theta,\omega)$ computed using    polynomials of degrees $2\leqslant p \leqslant 12$ and a reference cross-covariance $C_{\rm ref}(\Theta,\omega)$. Computations with and without flow are shown by the dashed and solid lines, respectively. The vertical line labeled `$p=10$' indicates the number of degrees of freedom used in all calculations in this paper.} \label{fig:accuracyTravelTimes} 
\end{figure}

\begin{table*}
  \centering
  \caption{Computational times and memory usage for single frequency runs of the radial (1.5D), axisymmetric (2.5D), and 3D methods for concentric shells with constant background coefficients. The number of modes (number of $l$ times number of $m$ for the radial case and number of $m$ for the axisymmetric case) is chosen such that the error is smaller than $10^{-4}$.
Computational costs are also given for a single mode run of the solar case in $1.5$D and $2.5$D.
All computational times are given for single core computations.}
  \label{table.CompData}
  \begin{tabular}{l l c c c} \hline\midrule
Simulation  &    & Radial & Axisymmetric & 3D \\ \midrule
 Concentric shells (Sect.~\ref{sec.concShells}) &&&& \\ 
& Number of degrees of freedom & $41$   & $4101$ & $453001$\\ 
& Number of modes              & $1369$   & $75$     & --- \\
& CPU time                     & $6.5$~s   & $42.4$~s  & $1753.2$~s\\
& Memory usage                 & $128$~MB & $173$~MB & $26.3$~GB \\ \midrule 
Solar model (Sect.~\ref{sec.TDH}) & &&&
  \\ 
& Number of degrees of freedom & $381$  & $1072900$ &  \\
& Number of modes              & $1$      & $1$         &  \\
& CPU time                     & $0.48$~s  & $141$~s   & \\
& Memory usage                 & $744$~kB & $6.6$~GB   & \\\midrule
  \end{tabular}
\end{table*}

\subsection{Convergence of travel times} \label{sect.convergenceTT}

Having discussed the accuracy of the Green's functions at different orders of discretization, we now examine the accuracy of the travel times. Based on Eq.~(\ref{eq:dtCross}), we compute the travel times for waves originating from the pole defined by:
\begin{equation}
\label{eq:def_tt_cicref}
\delta\tau^{(p)} = \int_{-\infty}^\infty W^*  (C^{(p)}-\CREF)\diff\omega.
\end{equation}
where $C^{(p)}$ is the cross-covariance computed from the Green's functions $G^{(p)}$ and $\CREF$ is the cross-covariance computed from the $\GREF$, as defined in the previous section. Green's functions were computed for a Nyquist frequency of $8.33$~mHz with a frequency resolution of $3.3$~$\mu$Hz and a constant damping rate of $\gamma/2\pi = 30~\mu$Hz. In Fig.~\ref{fig:accuracyTravelTimes}, we show that our method achieves a travel-time accuracy of 8~ms for the order of discretization $p=10$, with or without the presence of a background meridional flow, as in the previous section. This accuracy could be improved by a factor 10 if we used polynomials of degree $12$, however, the CPU time and memory requirements are increased by 100\% and 40\% respectively.

\subsection{Computational times for the $1.5$D, $2.5$D and $3$D problems}
\label{sec.comptimes}

Here we compare the computational times of the $2.5$D model with two other models (3D and radial $1.5$D). Initially, we examine a simple case of constant sound speed and density spherical layers with the Sommerfeld radiation boundary condition described in section~\ref{sec.concShells}. This is done in order to demonstrate the computational costs in all three model types. Following this, we compute the computational costs for the more demanding solar cases ($1.5$D and 2D) and neglect the full 3D case due to the high cost. However, we remind the reader that computational costs must be multiplied by the number of modes and frequencies required to accurately reconstruct the Green's function.

Table~\ref{table.CompData} shows CPU time and memory requirements of the three different models for a single frequency.  With each additional dimension, the required memory and CPU time become larger, with a dramatic increase in the full 3D case. 
A comparison of the model requirements shows that even though the axisymmetric method is more demanding than the radial one, the requirements are not unreasonable for most systems.

\section{Travel-time sensitivity kernels}

\begin{figure*}
\centering
\begin{minipage}{0.58\linewidth}
\begin{center}
\subfloat[]{\includegraphics[width=\linewidth]{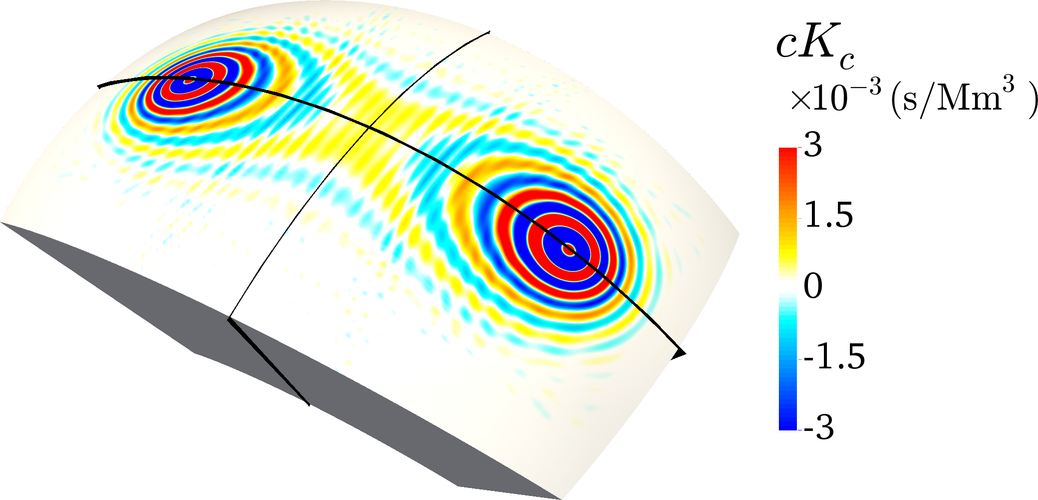}\label{fig.cKc3Dtop}}
\end{center}
\end{minipage}
\begin{minipage}{0.37\linewidth}
\begin{center}
\subfloat[]{\includegraphics[height=0.5\linewidth]{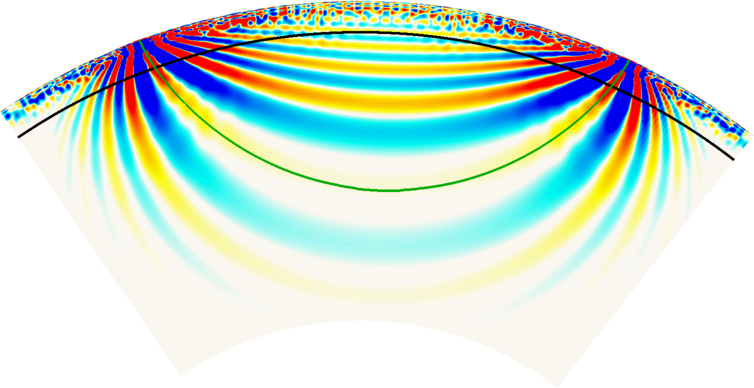}\label{fig.cKc3D_banana_small}}\\
\subfloat[]{\includegraphics[height=0.5\linewidth]{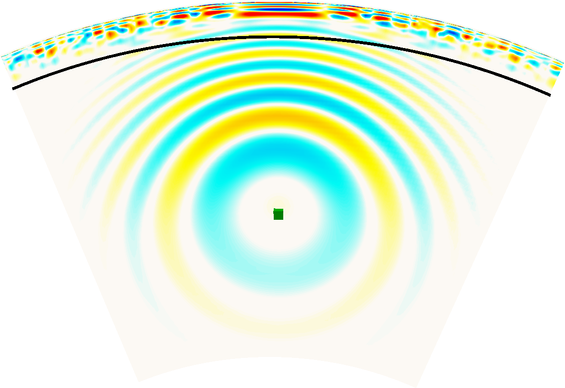}\label{fig.cKc3D_doughnut_small}}
\end{center}
\end{minipage}
\caption{
Sensitivity of the mean travel time, $[\tau(\one,\two)+\tau(\two,\one)]/2$,
to relative sound-speed perturbations $\delta c/c$ in the interior, with observation point $\one$ located on the polar axis and observation point $\two$ at latitude of $45^\circ$. Panel (a): Cut at $r=0.95R_\odot$ through the sound-speed kernel. 
Panel (b): Slice through a plane containing the two observation points and the center of the Sun. The green line indicates the position of the ray path.
Panel (c): Slice in a plane perpendicular to the ray path connecting the two observation points (green square), at equal distance from the two observation points.
In both slices (b) and (c), the black arc of a circle locates radius $r=0.95 R_\odot$  which corresponds to panel (a).}
\label{fig.Ku_3D}
\end{figure*}

In the results discussed thus far, we have focused on obtaining travel times through the direct modeling of waves propagating in an axisymmetric medium. In this section we address the computation and accuracy of the travel-time kernels outlined in section~\ref{sec.kernelForm}, which  describe the spatial sensitivity of travel times to local perturbations in the interior.

\subsection{Three-dimensional kernels}

Under the `convenient source' assumption, the computation of 3D kernels requires computing four Green's functions, as explained in Sect~6. When the medium does not contain a flow, only two Green's functions are needed. Even in this case, the computational burden is very demanding for a general three-dimensional background (Sect.~\ref{sec.comptimes}). However, it is feasible to compute 3D kernels when  the background model is axially- or spherically- symmetric, under the $2.5$D approach outlined thus far. 

In the case of spherical symmetry of the background, the kernels can be built from a single Green's function where the source is located on the pole at the observation height, after a series of rotations. Only the mode $m=0$ needs to be computed in this case.
This $m=0$ Green's function is then rotated to the desired source location ($\br_1$) and a duplicate is rotated to the receiver location ($\br_2$). The construction of the kernels  follows Eqs.~\eqref{eq.bilinearFormKernel}-\eqref{eq.bilinearFormKernel.end}.

Figures~\ref{fig.Ku_3D} shows slices through a sound-speed kernel with $\one$ at the pole and $\two$ at $45^\circ$ latitude, computed using the rotation of a $m=0$ Green's function and
$800$ frequencies between $1.5$~mHz and $4.5$~mHz. In this kernel we see the traditional banana-doughnut shape reported in geophysics \citep[Born-Fr\'echet kernels, e.g][]{marquering_etal_1999} and helioseismology \citep{Birch2004}. For the choice of observable that we have made, the travel-time sensitivity is very small near the ray path
(Fig.~\ref{fig.Ku_3D}c).  Surrounding the ray path are regions of negative and positive sensitivities corresponding to the consecutive Fresnel zones. The values near the surface are not numerical noise but are due to high spatial frequencies. Fig.~\ref{fig.Ku_3D} is a 3D illustration of the sound-speed kernel that also shows the values on the sphere at radius $r=0.95 R_\odot$. The sensitivity is maximum near the surface around the two observation points. Notice that we have plotted the product $c K_c$ to better render the deeper layers. The computation of the $m=0$ Green's function took approximately $1$~hr using $200$~cores. The post-processing consists mostly in computing the rotated Green's functions and took approximately $3$~hr.

\subsection{Longitudinally averaged  kernels}

\begin{figure*}
\includegraphics[width=\linewidth]{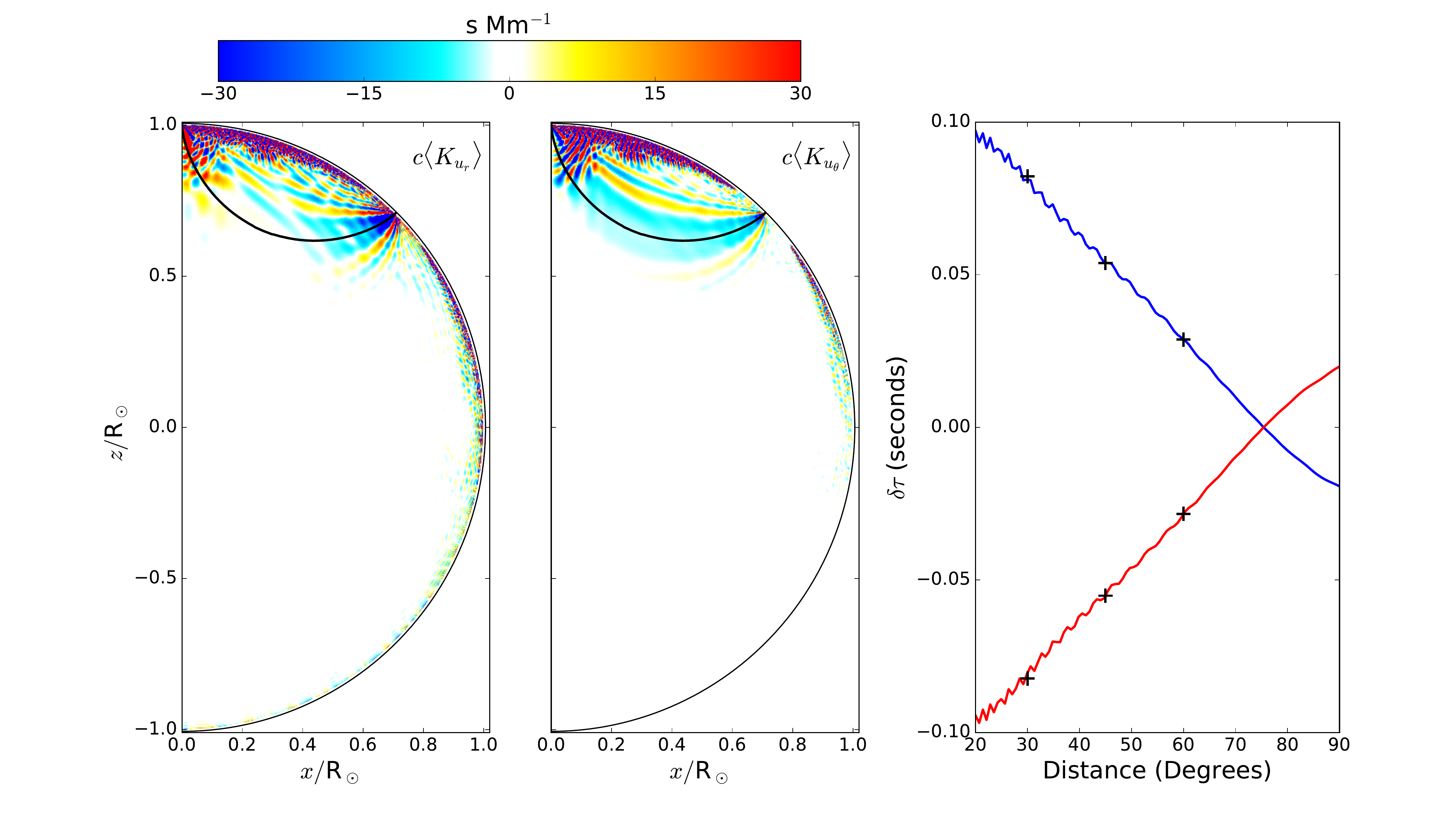}
  \caption{{Left and center panels: 
  Kernels $\langle K_{u_r} \rangle$ and $\langle K_{u_\theta} \rangle$ for the  $r$ and $\theta$ components of the flow.
  Point $\one$ is at the north pole (photosphere) and
  point $\two$ is at $45^\circ$ latitude. The values of the kernels are scaled by the sound speed $c$ and are saturated at $1/400$-th of the maximum value. The ray-path connecting the two points is shown (thick black line) as well as the computational boundary (black half circle). Right panel: 
  {Comparison between the travel times computed directly from the cross-covariance function (curves, see Eq.~\eqref{eq:dtCross}) and those computed from the sensitivity kernels (`+' symbols, see Eq.~\eqref{eq.longazidt}).} The blue curve is for the travel times measured from the pole to latitude $45^\circ$, the red curve for the travel times measured in the opposite direction. The accuracy of the travel times is of order $10^{-3}$~s.}}
  \label{fig.m0_kernel}
\end{figure*}

For an axisymmetric (but not necessary spherically symmetric) background, the Green's functions must be constructed by summing a sufficient number of $m$ modes. 
Here, in addition, we consider axisymmetric perturbations only, $q_\alpha=q_\alpha(r,\theta)$, and we wish to determine the 2D spatial sensitivity of the travel times by averaging the kernels over longitude. Useful applications include rotation ($u_\phi$) and meridional circulation ($u_\theta$).
For axisymmetric perturbations, we have
\begin{equation}\label{eq.longazidt}
\delta \tau(\one,\two) = 2\pi \sum_\alpha  \int_0^\pi \id\theta \int_0^R \id r \; \delta q_\alpha(r,\theta) \;
  \langle{K}_\alpha\rangle(r,\theta; \br_1, \br_2) 
\end{equation}
where $\langle{K}_\alpha\rangle$ is the longitudinally averaged kernel 
\begin{equation}
\langle {K}_\alpha\rangle(r,\theta; \br_1, \br_2)  
=
\frac{r^2 \sin \theta}{2\pi}  \int_0^{2\pi}  K_\alpha (r,\theta, \phi; \br_1, \br_2) \, \id\phi.
\end{equation}
Let us define the Green's function and the cross-covariance function in terms of their  longitudinal mode components $G^m$ and $C^m$ as follows:
\begin{equation}
\begin{aligned}
& G (\br, \br_1,\omega )  = \sum_{m=-\infty}^\infty G^m(\tilde{\br}, \tilde{\br}_1,\omega) \, \mathrm{e}^{\ii m  ( \phi-\phi_1)}, 
\\
& \overline{C} (\br_1, \br,\omega )  = \sum_{m=-\infty}^\infty  C^m(\tilde{\br}_1, \tilde{\br}, \omega) \, \mathrm{e}^{\ii m  ( \phi-\phi_1)} .
\end{aligned}
\end{equation}
Notice the order of the variables in $G$ and $\overline{C}$ in the above notations.
Using the formulation of the kernels given by Eq.~\eqref{eq.kernelGeneral}, we obtain
\begin{equation}
\begin{aligned}
 &K_\alpha (r,\theta , \phi; \br_1, \br_2) =  \\
  & - \int_{-\infty}^\infty \id\omega W^* \sum_{m, m'} \cL_\alpha \left[ G^m(\tilde{\br}_2, \tilde{\br} ) \mathrm{e}^{\ii m(\phi_2-\phi)} 
, {C}^{m'}(\tilde{\br}_1, \tilde{\br} ) \mathrm{e}^{-\ii m'(\phi_1-\phi)} 
\right]  \\ 
& - \int_{-\infty}^\infty \id\omega W^* \sum_{m, m'} \cL_\alpha^* \left[ G^{m*}(\tilde{\br}_1, \tilde{\br} ) \mathrm{e}^{-\ii m(\phi_1-\phi)} 
, {C}^{m'*}(\tilde{\br}_2, \tilde{\br} ) \mathrm{e}^{\ii m'(\phi_2-\phi)} 
\right].
\end{aligned}
\end{equation}
Using the explicit expressions for the bilinear operators $\cL_\alpha$ (Eqs.~\eqref{eq.bilinearFormKernel}-\eqref{eq.bilinearFormKernel.end}), we can then  obtain the longitudinally averaged kernels for all perturbations $q_\alpha$.

As an example, the flow kernels can be written as a sum 
\begin{equation}\label{eq.mKernel}
\langle K_{u_k}\rangle 
(r, \theta; \br_1, \br_2) 
= \sum_{m=-\infty}^{\infty}  \langle K_{u_k}\rangle^m 
(r, \theta;\br_1, \br_2) 
\end{equation}
over azimuthal components:
\begin{equation}\label{eq.flowmKernel}
\begin{aligned}
\langle K_{u_k}\rangle^m &= 2\ii  \rho r^2 \sin\theta \int_{-\infty}^\infty \id\omega \, \omega W^\ast \mathrm{e}^{\ii m (\phi_2 - \phi_1)} 
\\
&  \!\!\!\!\!\! \times \left[ G^m(\tilde{\br}_2,\tilde{\br}) \, 
\tilde{\partial}_k  C^{-m}(\tilde{\br}_1,\tilde{\br}) 
-G^{m*}(\tilde{\br}_1, \tilde{\br}) \,
\tilde{\partial}_k C^{-m*}(\tilde{\br}_2, \tilde{\br}) \right] ,
\end{aligned}
\end{equation}
where the operator $\tilde{\partial}_k$ is either
$\tilde{\partial}_r=\partial_r$, $\tilde{\partial}_\theta= \partial_\theta/r$, or $\tilde{\partial}_\phi=\ii m / (r \sin\theta)$.

In practice, $G^m(\tilde{\br}_j, \tilde{\br},\omega)$ is obtained using generalized seismic reciprocity, $G^m(\tilde{\br}_j, \tilde{\br},\omega)=G^{-m}(\tilde{\br},\tilde{\br}_j,\omega; -\bu)$, performing a simulation with a source at $\tilde{\br}_j$. Using the convenient source of excitation, the cross-covariance is linked to the Green's function by Eq.~\eqref{eq.EC-ImG}. One can obtain a similar relation for the Fourier coefficients:
\begin{equation}
 C^m (\tilde{\br}_1,\tilde{\br}, \omega)  = \frac{\Ps(\omega)}{4 \ii \omega}
 \left[ G^m (\tilde{\br}, \tilde{\br}_1, \omega) - G^{-m\ast}(\tilde{\br}, \tilde{\br}_1, \omega; -\bu) \right] . \label{eq.Cm}
\end{equation}
This means that the (3D) kernels can be computed using only the azimuthal modes of the Green's function obtained from the $2.5$D solver. 

Several comments can be made:
\begin{itemize}
\item If one of the observation points is on the rotation axis, then only the $m=0$ mode of the Green's function is required.  For a measurement between two arbitrary points at the surface of the Sun, the computation of many modes is required (see next subsection). 

\item If the background model contains a flow, then the Green's function for a reversed flow is also required in order to compute the cross-covariance as shown by Eq.~\eqref{eq.Cm}. 

\item Eq.~\eqref{eq.wavem} shows that, in the case of no background flow, only the $m\ge 0$ solutions need to be computed since $G^{-m}=G^m$. 
\item If the kernels for points $(\theta_1, \phi_1)$ and $(\theta_2, \phi_2)$ have been computed, then we obtain for free the kernels for points at the same latitudes but any longitudes $\phi_1$ and $\phi_2$ since the only term depending on longitude is the exponential ${\rm e}^{\ii m(\phi_2-\phi_1)}$ in Eq.~\eqref{eq.flowmKernel}.
\end{itemize}

\subsection{Accuracy of travel-time kernels for flows}

\begin{figure*}[!htb]
\includegraphics[width=\linewidth]{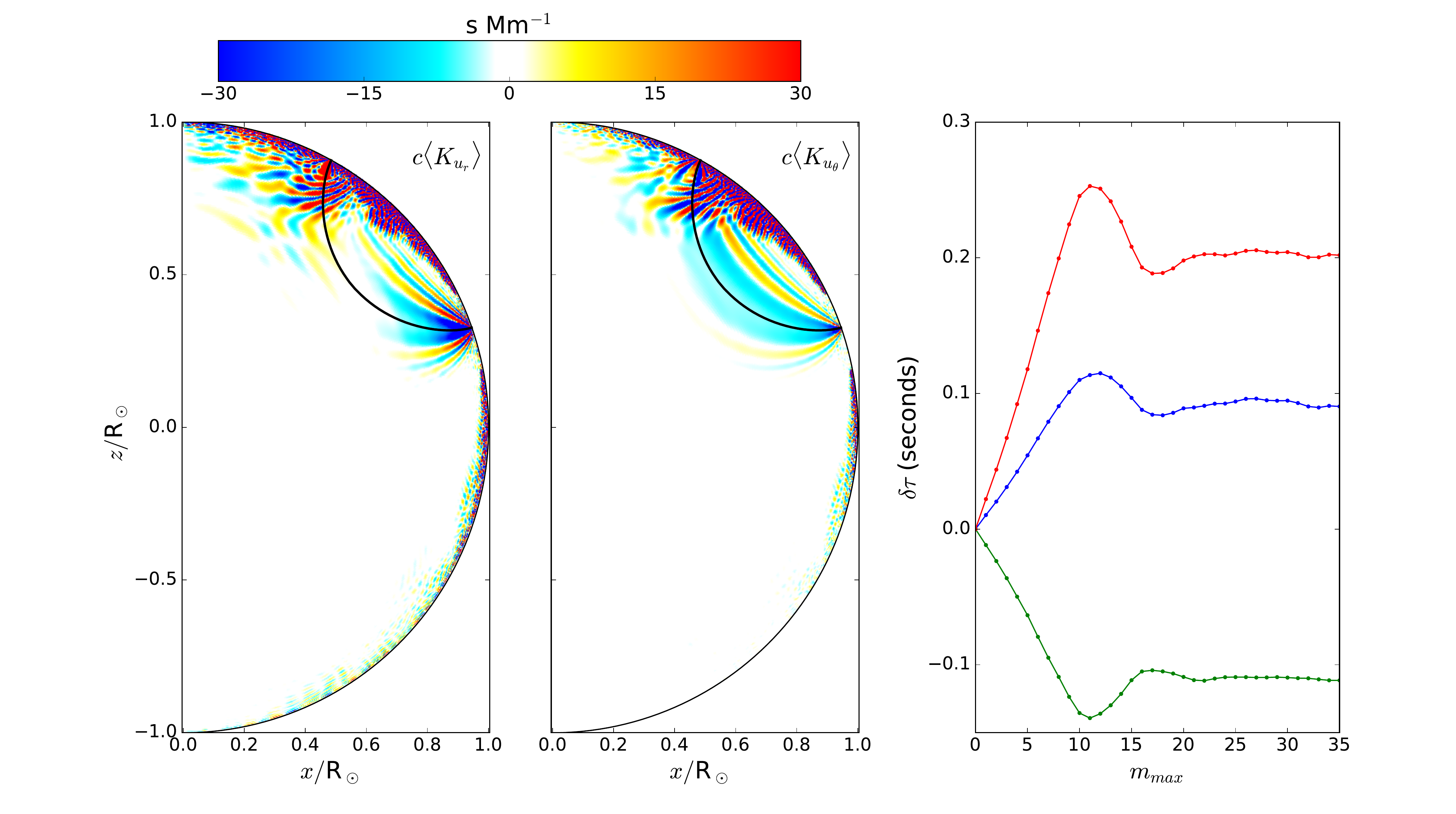}
  \caption{
  Left and center panels: 
  Kernels $\langle K_{u_r} \rangle$ and $\langle K_{u_\theta} \rangle$ for the  $r$ and $\theta$ components of the flow, using all azimuthal components $|m|\leq m_{\rm max}=35$.
 The separation distance between points $\one$ and $\two$ is $42^\circ$ with the center point located at a latitude of $40^\circ$.
 The ray path connecting the two points is shown in black. The right panel shows the convergence of the travel times as a function of $m_{\rm max}$, for the North-South (blue) and South-North (green) travel directions. The red curve shows their difference.}
  \label{fig.m_kernel}
\end{figure*}

Consider a spherically symmetric background reference model with no flow.  Figure~\ref{fig.m0_kernel} shows the flow kernels $\langle{K}_{u_r}\rangle = \langle{K}_{u_r}\rangle^{m=0}$ and $\langle{K}_{u_\theta}\rangle= \langle{K}_{u_\theta}\rangle^{m=0}$ with a point $\one$ on the pole (at the solar surface) and a point $\two$ located at latitude $45^\circ$. The kernel $\langle{K}_{u_\phi}\rangle^{m=0}$ is zero
by construction because $\langle{K}_{u_\phi}\rangle^{m}$ is proportional to $m$. 
{The 2D kernels for $u_r$ and $u_\theta$ display Fresnel zones as do the 3D kernels, however, the null points along the ray path are absent here because of integration in the longitudinal direction. }
Additionally, as reported by \citet{Birch2007}, the kernels exhibit hyperbola-like patterns near point $\one$. This pattern is due to scattering from distant  sources \citep{GB2002} and is not present in Earth seismology kernels for point-source earthquakes. 
We refer the reader to the work by \citet{Duvall2006} for an observational study of 2D horizontal sensitivity kernels.
The right-hand panel of Fig.~\ref{fig.m0_kernel} compares the travel times computed in two different ways to evaluate the accuracy of the kernels: (1) by multiplying the kernels by a flow model $\bu$ and integrating (Eq.~\eqref{eq.longazidt}) and (2) by computing the difference $\delta C = C(\bu) - C(\bu=0)$ and then measuring the travel time (Eq.~\eqref{eq:dtCross}).
Here we have used the meridional flow model shown in Appendix~\ref{app:flowCell} with a maximum flow speed of $20$ms$^{-1}$ at the surface. 
For the comparison we have considered 
three separation distances ($30^\circ$, $45^\circ$, and $60^\circ$) and the two directions, $\delta\tau(\one, \two)$ and $\delta\tau(\two,\one)$. 
The kernel-based computation of the travel times and the direct computation agree to within $10^{-3}$~s.
This is an important result as it demonstrates that our kernels have sufficient accuracy for the interpretation of solar travel times.
For meridional circulation measurements, the noise in the travel times is typically of order $0.1$-$0.5$~s by averaging data over four years \citep{rajaguru2015}.

{For practical applications, the two points should be off-axis since helioseismic observations are limited to a center-to-limb distance of $\sim 70^\circ$. In such cases the azimuthally averaged kernels require the computation of many components $\langle K \rangle^m$. Figure~\ref{fig.m_kernel} shows the components of flow kernel computed from Eq.~\eqref{eq.flowmKernel} for all $m \leq 35$. In this figure the travel times are measured between points at latitudes $61^\circ$ and $19^\circ$, both along the central meridian. With this separation distance of $42^\circ$ the ray path reaches a depth of $0.72~R_\odot$. Like before, the kernels are not symmetric about the center point between the observation points and have features similar those seen in Fig.~\ref{fig.m0_kernel}. }

{To test the convergence of these kernels, we calculate individual kernels including all modes $|m|\leq m_{\rm max}$ and compute  travel times as a function of $m_{\rm max}$ in the presence of the same meridional flow as used previously. The right-hand panel of Fig.~\ref{fig.m_kernel} shows that the travel times converge to an asymptotic value for $m_{\rm max}>25$  with an accuracy of $\sim 0.01$~s. We note that a larger $m_{\rm max}$ is needed to achieve convergence for shorter  separation distances.}

\subsection{Filtering}

Filtering the observations in the $\ell$-$\omega$ domain is common practice in helioseismology. Several choices of filters have been proposed: filters in $\omega$ space, phase-speed filters, and ridge filters \citep[see, e.g.,][]{GB2005}. To interpret any particular travel-time measurement, the sensitivity kernels must account for the proper frequency content of the seismic data set by the filtering. This dependency of the kernel on the filter has been studied previously \citep[e.g.,][]{Birch2004,Boening2016}.

\begin{figure*}[!htb]
\includegraphics[width=\linewidth]{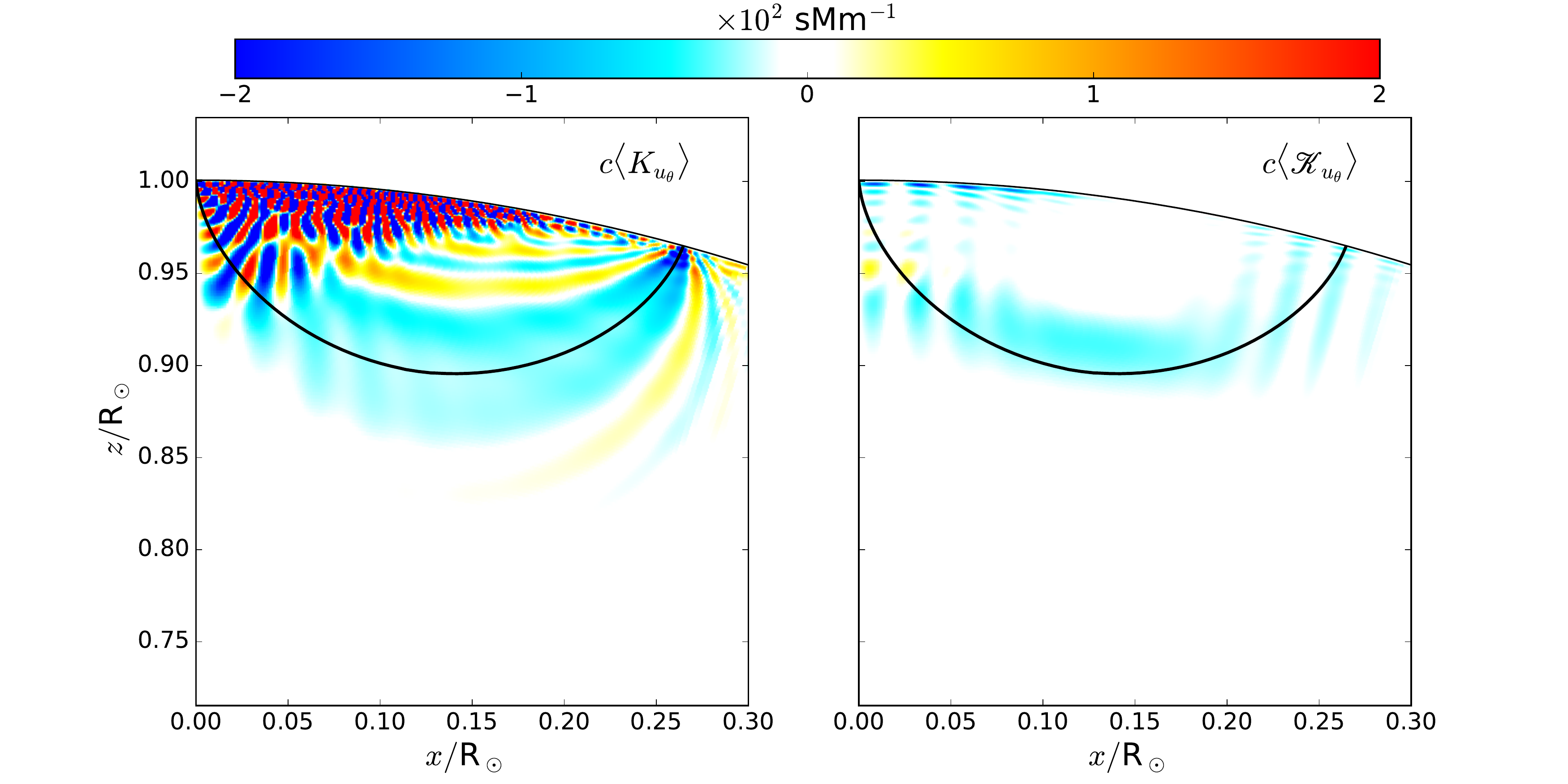}
  \caption{Left: Unfiltered $\langle K_{u_\theta}\rangle$ kernel with $\one$ at the pole and $\two$ at a co-latitude of $15.36^\circ$. Right: Kernel $\langle \mathscr{K}_{u_\theta}\rangle$ for filtered observations, where the Gaussian phase-speed filter $F_\ell(\omega)$ is centered at $125.2$~km/s with a dispersion of $12.3$~km/s, for $\ell$ up to 1000. The ray path is shown by the black line.}
  \label{fig.m_kernel_Filtered}
\end{figure*}

Symbolically, the filtered observation, $\Psi$, is obtained by 
applying a filtering operator, $\mathscr{F}$, to the original wavefield, $\psi$:
\begin{equation}
\Psi(\br, \omega) = \mathscr{F}[\psi(\br,\omega)]    .
\end{equation}
The corresponding kernel is obtained by applying the filtering operator twice to the original kernel:
\begin{eqnarray}
\mathscr{K}_\alpha(\br, \br_1, \br_2) &=&
  - \int_{-\infty}^\infty W^* \cL_\alpha\left[ \mathscr{G}(\br_2, \br, \omega),  {\mathscr{C}}(\br_1,\br, \omega) \right]
\id\omega \nonumber
\\
&& -
 \int_{-\infty}^\infty W^* \cL_\alpha^*\left[ \mathscr{G}^*(\br_1, \br, \omega),  {\mathscr{C}}^*(\br_2,\br, \omega) \right]
\id\omega ,  \nonumber\\
\end{eqnarray}
where the filtered Green's function and cross-covariance are
\begin{eqnarray}
\mathscr{G}(\br_1,\br,\omega) &=& \mathscr{F}_1[G(\br_1,\br,\omega)] , \\
{\mathscr{C}}(\br_1,\br,\omega) &=& \mathscr{F}_1 [ \overline{C}(\br_1,\br,\omega)] \nonumber \\
&=& \frac{\Ps(\omega)}{4\ii \omega} \left[ \mathscr{G}(\br_1,\br,\omega; -\bu) - \mathscr{G}^*(\br_1,\br,\omega) \right],  
\end{eqnarray}
and the $W$ function is computed with the choice $C_{\rm ref}=\mathscr{C}$. Here, $\mathscr{F}_1$ indicates that the filtering has to be done with respect to the point $\br_1$.

The filtered Green's function may be obtained by filtering the delta source function. To see this, we use generalized seismic reciprocity:
\begin{eqnarray}
\cG(\br_1, \br, \omega) &= & \mathscr{F}_1 [G(\br, \br_1, \omega; -\bu )] \nonumber \\
&= & \int_V G(\br, \br', \omega; -\bu) \,  \mathscr{F}_1[ \delta(\br_1-\br')] \,\id\br'  ,
\end{eqnarray}
where 
\begin{eqnarray}
\cF_1 [\delta(\br_1-\br') ]
& = & \delta(r_1-r') \; \mathscr{F}_1 \left[ \sum_{\ell = 0}^\infty \sum_{m=-\ell}^\ell Y_\ell^m(\rhat_1) Y_\ell^{m*}(\rhat') \right]
\nonumber \\
& =&  \delta(r_1-r')  \sum_{\ell = 0}^\infty F_\ell(\omega) \frac{2 \ell+1}{4\pi} P_\ell(\rhat_1\cdot\rhat') \label{eq:filteredS}
\end{eqnarray}
is the filtered delta function source. In the above expression   $F_\ell(\omega)$ can be either a phase-speed filter or a ridge filter.  Note that the filtered source function is a function of frequency.

If the background is spherically symmetric with no background flow, it is also possible to compute the non-filtered Green's function and to perform the filtering a posteriori. In this case, the Green's function $G(\br_1, \br, \omega)$ depends only on the depths $r_1$ and $r$ and on the angular distance $\Theta_1$ between the two points
\begin{equation}
G(\br_1, \br, \omega) = G(r_1, r, \Theta_1, \omega),
\end{equation}
where $\cos\Theta_1 = \rhat_1\cdot\rhat$.
The filtered Green's function takes the form
\begin{equation}
\begin{aligned}
\mathscr{G}(\br_1,\br,\omega) = &   \sum_{\ell=-\infty}^\infty F_\ell(\omega)    G_{\ell}(r_1,  r, \omega) P_\ell(\cos\Theta_1) , \label{eq:filteredG}
\end{aligned}
\end{equation}
where
\begin{equation}
 G_{\ell}(r_1,r,\omega) = \frac{2\ell+1}{2} \int_0^\pi  G(r_1, r,\Theta,\omega) P_\ell(\cos\Theta) \sin \Theta \diff \Theta \label{eq:Gl}
\end{equation}
is the projection of $G$ on the Legendre polynomials. Using seismic reciprocity, we obtain $G_{\ell}(r_1,r,\omega) = G_{\ell}(r,r_1,\omega)$ and Eq.~\eqref{eq:filteredG} implies
\begin{equation}
\mathscr{G}(\br_1,\br,\omega) = \mathscr{G}(\br,\br_1,\omega).
\end{equation}
Thus, if the background is spherically symmetric, we also have seismic reciprocity for the filtered Green's function: the filtering can be seen as a post-processing operation on $\br$ at fixed source position $\one$. Otherwise, one should use a filtered source as given by Eq.~\eqref{eq:filteredS}.

As a special case, we computed the azimuthally averaged kernels of Eq.~\eqref{eq.flowmKernel} for filtered observations when $\one$ is located at the pole and $\bu=\mathbf{0}$. As mentioned previously, only the $m=0$ component of the Green's function is needed. {Figure~\ref{fig.m_kernel_Filtered} shows the effect of applying a phase-speed filter to the kernel $\langle K_{u_\theta}\rangle$ with $\one$ at the pole and $\two$ at a co-latitude of $15.36^\circ$. The  phase-speed filter is centered at $125.2$~km/s with a width of $12.3$~km/s. These values were chosen to be the same as for filter \#11 of  \citet{duvall_hanasoge_2013}.  Once the filter is applied the sensitivity to $u_\theta$ is predominantly near the base of the ray-path. This filtered  kernel appears to be similar to the $K_6$ case from \citet{Boening2016} indicating general agreement with their work.}

\section{Conclusion}

{
We have presented a new framework for computational helioseismology by solving the forward problem in the frequency domain. For the sake of simplicity, we considered a simplified scalar acoustic wave equation and assumed spatially uncorrelated sources of excitation distributed through the Sun. Under such conditions, the cross-covariance can be obtained directly from the imaginary part of the frequency-domain Green's function. 
{This leads to a  convenient, flexible and fast  way to compute accurate kernels.} The analytical work involved in this framework is less cumbersome than in previous work that relies on normal-mode expansions of the kernels \citep[e.g.,][]{Birch2007,Burston2015, Boening2016}. The framework can relatively easily be extended to the vectorial wave equation using existing Montjoie vectorial setups \citep[e.g.][for Maxwell's equations]{durufle_MAXWELL}. The scalar equation captures the propagation of acoustic waves through a solar-like medium and leads to an oscillation power spectrum that compares well with observations. The present setup will be very useful to test new methods, include instrumental and projection effects, but also to interpret existing travel-time measurements for rotation, meridional circulation, and axisymmetric structures like the average supergranule.   


{In future work we intend to address the inverse problem. Specifically, we wish to find the parameters $\delta q$ of the background model (sound speed, density, flows) such that the travel times $\tau$ from the model are consistent with the observed travel times  $\tau_{\rm obs}$. This is generally done by solving a linear system of the form $\delta\tau  =  \text{K}  \, \delta q    + n$, where K is a matrix of kernels and $n$ is a vector of travel-time noises, defined through the noise covariance matrix $\Lambda=\EE[n\, n^T]$ \citep[see][]{GB2004,Fournier2014}. This problem can  be solved by classical regularization methods such as Regularized Least Square (RLS) \citep[e.g.,][]{KOS96} or Optimally Localized Averaging (OLA) \citep[e.g.,][]{HAB04}.
Another approach is to solve a nonlinear inverse problem defined in terms of the partial differential equation. Different methods exist such as Landweber iteration  \citep{HNS:95}, Newton type methods such as the iteratively regularized Gauss-Newton \citep{BAK92} or Newton Conjugate Gradient methods \citep{hanke:97}. 
Most of these methods avoid the explicit computation of 
sensitivity kernels.  All these inverse methods are feasible under the assumption of an axisymmetric background model, thanks to the embarassingly parallel workload in $m$ and $\omega$. The last iteration of the inversion produces a three-dimensional model of the solar internal properties. Using a full three-dimensional forward solver is currently not practical. }

\begin{acknowledgements}
LG and ACB developed the general theoretical framework. HB, JC, and MD developed tools to solve the wave equation (Montjoie package). CSH, ML, and DF implemented the tools in the solar context, tuned the power spectrum, and computed the kernels. All authors contributed to the writing of the paper. LG acknowledges generous financial support from  the State of Lower Saxony, Germany, and partial support from the Center for Space Science at the NYU Abu Dhabi Institute, UAE, under grant G1502. The computer infrastructure was provided by the German Data Center for SDO funded by the German Aerospace Center (DLR). The Global Oscillation Network Group is funded by the National Science Foundation. The HMI project is supported by NASA contract NAS5-02139.

\end{acknowledgements}

\bibliographystyle{aa} 
\bibliography{aa} 

\appendix

\section{Solving the eigenvalue problem using ADIPLS}
\label{app:ADIPLS}
\newcommand{\TD}[2]{\frac{\mathrm{d}{#1}}{\mathrm{d}{#2}}}
\newcommand{\rd}{\mathrm{d}}

We modified ADIPLS to solve the eigenvalue problem 
\begin{equation}
\cH [\bxi]= \omega^2\bxi
\end{equation}
where $\cH$ is given by Eq. (\ref{eq.Hmaster}) and we neglected the gravity terms.
Using the same notation as the one used by \citet{Christensen-Dalsgaard2008}, 
the corresponding eigenvalue problem for the eigenfrequency and the eigenfunction of a mode takes the form 
\begin{eqnarray}
 \omega^2 \xi_r & = & 
   - \frac{1}{\rho} \TD{}{r}\left[ \rho c^2 \left(
      \TD{\xi_r}{r} + 2 \frac{\xi_r}{r} - L^2 \frac{\xi_h}{r}
      \right ) \right ] , \\
\omega^2\xi_h & = & -
       c^2 \left ( \frac{1}{r}\TD{\xi_r}{r}  +2\frac{\xi_r}{r^2} -L^2\frac{\xi_h}{r^2}\right ) ,
\end{eqnarray}
where $L^2=\ell(\ell+1)$, and $\xi_r(r)$ and $\xi_h(r)$ are the radial and horizontal eigenfunctions. The equations can be rewritten as
\begin{eqnarray}
r\TD{\xi_r}{r} & = & -2\xi_r + \left ( L^2 - \frac{\omega^2 r^2}{c^2} \right ) \xi_h, \\
r\TD{\xi_h}{r} & = & \xi_r - \left ( 1+ \frac{\rd \ln \rho }{\rd \ln r} \right ) \xi_h,
\end{eqnarray}
which, in ADIPLS adimensionalized form \citep[see ][]{Christensen-Dalsgaard2008}, become
\begin{eqnarray}
\label{eq:adi1}
x\TD{y_1}{x} & = & -2 y_1 + \left ( 1 - \frac{V_g}{\eta} \right ) y_2, \\
\label{eq:adi2}
x\TD{y_2}{x} & = & L^2 y_1 + \left ( A + V_g - 1 \right ) y_2.
\end{eqnarray}
We modified ADIPLS, by changing the file \texttt{rhs.n.d.f} accordingly in such a way that it solves Eqs. (\ref{eq:adi1}) and (\ref{eq:adi2}), with a free surface boundary condition. 

We calculated the frequency splittings in presence of differential rotation by evaluating the integral
\begin{equation}
\delta\omega_{n\ell m} = m \int_0^R \id r \int_0^\pi K_{n\ell m}(r,\theta) \Omega(r,\theta) \, r \rd{\theta}
\end{equation}
where the rotation profile $\Omega(r,\theta)$ is defined  by Eq. (\ref{eq.rotprofile}), and  $K_{n\ell m}(r,\theta)$ are the rotational sensitivity kernels \citep[see Chapter 8.4 of][]{JCD}, which depend on $\xi_r$ and $\xi_h$.

\section{A meridional flow model}
\label{app:flowCell}

\newcommand{\StreamF}{\mathbf{\Psi}}
\newcommand{\RTOP}{r_\mathrm{t}}
\newcommand{\RBOT}{r_\mathrm{b}}
\newcommand{\UTOP}{U}

Various models of meridional flow cells can be found in the literature \citep[e.g.][]{van_Ballegooijen_choudhuri_1988,dikpati_choudhuri_1995,jouve_etal_2008}.
In these models the flow is expressed through a stream function
$\Psi$ as follows:
\begin{equation}
  \bu(r,\theta) = \dfrac{1}{\rho}\nabla\times(\Psi \phihat) .
\end{equation}
This enforces mass conservation, $\nabla\cdot(\rho \bu) = 0$.
However, we cannot use directly the various expressions of the meridional flow from previous work, as they were not computed with the density $\rho(r)$ of model S.

\begin{figure}
  \centering
  \includegraphics[width=0.8\linewidth]{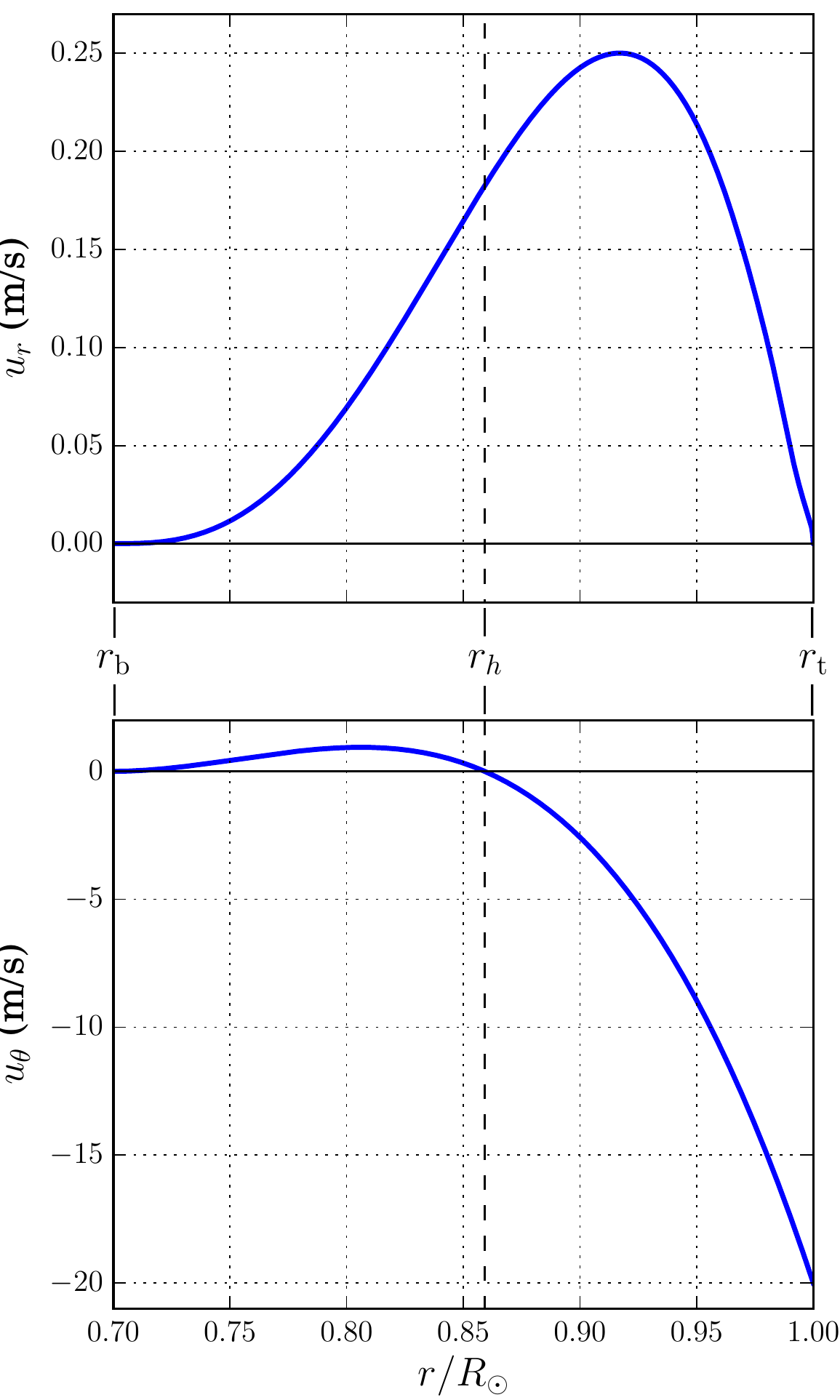}
  \caption{Plot of $u_r(r,\theta)$ (left) and $u_\theta(r,\theta)$ (right) at $\theta=45^\circ$ of the meridional flow model discussed in the appendix for a maximum flow velocity $U=20$~m/s at the surface.}
  \label{fig:flowProfiles}
\end{figure}

Let us define a stream function $\Psi(r,\theta)$ for $r_b \leqslant r \leqslant r_t$ and $0\leqslant \theta\leqslant \pi$, where $r_b$ and $r_t$ refer to the bottom and the top of the convection zone respectively. $\Psi$ is set to zero outside this region. 
Let us look for a solution of the form
\begin{equation}\label{eq:streamFunc}
  \Psi(r,\theta) = \rho_t r_t U   f(r)g(\theta),
\end{equation}
where  $f(r)$ and $g(\theta)$ are dimensionless functions to be determined, $U$ sets the amplitude of the flow, and $\rho_t=\rho(r_t)$ is the density at the top of the cell. The flow is given by
\begin{eqnarray}
  u_r(r,\theta) & = & U \frac{\rho_t r_t }{\rho r}  \frac{f(r)}{\sin\theta}  \frac{{\rm d}}{{\rm d} \theta} [g(\theta) \sin\theta ],
  \\
 u_\theta(r,\theta) & = & -U \frac{\rho_t  r_t}{\rho r} g(\theta) \frac{{\rm d}}{{\rm d} r} [rf(r)].
\end{eqnarray}
For one cell per hemisphere with $r_b<r<r_t$, the functions $f$ and $g$ satisfy the following conditions:
\begin{equation}
  f(r_b) = f(r_t) = 0
  \text{ and }
  g(0) = g(\pi) = 0.
\end{equation}
For the latitudinal dependence, we choose 
\begin{equation}
g(\theta)=\sin(2\theta).
\end{equation}
We seek the function $f$ in terms of the function 
\begin{equation}\label{eq:radialProfileUTh}
  h(r) = -\frac{\rho_t r_t}{\rho r}\frac{{\rm d}}{{\rm d} r}( rf ),
\end{equation}
which controls the radial profile of $\rho u_\theta$. We have
\begin{equation}\label{eq:intFlowF}
  f(r) = -\frac{1}{\rho_t r_t r}\int_{r_b}^{r}  h(r')   \rho(r')  r' \, \diff r'.
\end{equation}
The description of the flow cell then relies on the choice of the function $h(r)$. To proceed, we choose the following conditions:
\begin{equation}
  h(r_t) = 1,\ h(r_b) = 0,\ \frac{{\rm d}h}{{\rm d}r}(r_b)=0 
\end{equation}
While the first two conditions are intuitive, the third one is arbitrary (other choices would be possible). 
We choose $h$ as a fourth order polynomial:
\begin{equation}
  h(r) =  (r-r_b)^2(r-r_h)(r-a) / r^4_t,
\end{equation}
where $a = r_t - r_t^4 (r_t-r_b)^{-2} (r_t-r_h)^{-1}$ is implied by $h(r_t)=1$.
In this paper, we set $r_b=0.7 R_\odot$ and $r_t = R_\odot$.
The depth $r_h$ at which the horizontal flow switches sign is such that $f(r_h)$ = 0. The value of $r_h$ can be obtained by interpolation or by a Newton method. We find $r_h = 0.859 R_\odot$. For the amplitude of the flow, we take  $U=20$~m/s.
The radial and co-latitudinal flows, given by 
\begin{eqnarray}
  u_r(r,\theta) & = & U \frac{\rho_t r_t }{\rho r}  \, f(r) \, 2[\cos^2\theta + \cos(2\theta)]  ,  \\
 u_\theta(r,\theta) & = &  U \, h(r)   \, \sin(2\theta) ,
\end{eqnarray}
are plotted as a function of $r$ at co-latitude $\theta=45^\circ$ in Fig.~\ref{fig:flowProfiles}.

\end{document}